\documentclass[a4paper,11pt]{article}
\usepackage{a4wide,amsmath,amssymb,slashed,bbm,bm}
\usepackage[english]{babel}

\usepackage{cite}

\makeatletter
\g@addto@macro\bfseries{\boldmath}

\@addtoreset{equation}{section}
\makeatother

\begin{document}

\hfill{ %Imperial-TP-LW-2016-04
}

\vspace{30pt}

\begin{center}
{\Huge{\bf All symmetric $AdS_{n>2}$ solutions \\ of type II supergravity}}

\vspace{50pt}

{\bf Linus Wulff}

\vspace{15pt}

{\it\small Department of Theoretical Physics and Astrophysics, Masaryk University, 611 37 Brno, Czech Republic}\\

\vspace{100pt}

{\bf Abstract}
\end{center}
\noindent
All symmetric space $AdS_n$ solutions of ten-dimensional type IIA and type IIB supergravity are constructed for $n>2$. There are a total of 26 geometries. Out of these only 7 allow for supersymmetries and these are the well known backgrounds coming from near-horizon limits of (intersecting) branes in ten or eleven dimensions and preserving 32, 16 or 8 supersymmetries.

\pagebreak 
\tableofcontents

\setcounter{page}{1}

%%%%%%%%%%%%%%%%%%%%%%%%%%%%%%%%%%%%%%%%%%%%%%%%%%%%%%%%%%%%%%%%%%%%%%%%

\section{Introduction}
Symmetric spaces provide the simplest class of supergravity solutions since for these the supergravity equations become algebraic. By a symmetric supergravity solution we mean that the geometry is that of a symmetric space and also that the fluxes (and dilaton) are expressed in terms of invariant forms on this symmetric space, i.e. they respect the isometries of the space. Since the invariant forms must be both constant and covariantly constant all terms in the supergravity equations involving derivatives of the fluxes or dilaton drop out. Furthermore the Ricci tensor for a (irreducible) symmetric space is proportional to the metric and the supergravity equations reduce to simple quadratic equations relating the parameters in the fluxes to the curvature radii of the geometry. Solving these quadratic equations, although in principle straight-forward, can be quite {involved (and the set of solutions very rich) if the space in question allows for many invariant forms.} In the case of eleven-dimensional supergravity this was carried out in \cite{Wulff:2016vqy}, completing earlier work in \cite{FigueroaO'Farrill:2011fj}.

Here we want to attempt the same classification in ten dimensions. Specifically we will consider the maximal type IIA and type IIB supergravities. {A partial classification for the type IIB case was obtained in \cite{FigueroaO'Farrill:2012rf}, we comment below on the overlap with the results presented here.} Although the number of allowed geometries goes down in going from eleven to ten dimensions the complexity of the analysis goes up. The reason for this is that, while in eleven dimensions there is only the four-form flux, in ten dimensions there are RR fluxes of various degree as well as the three-form NSNS flux. There are therefore many more parameters that enter the ansatz for the fluxes. In particular for $AdS_2$ solutions this makes the analysis so involved that we will not attempt it here (some type IIB solutions were found in \cite{FigueroaO'Farrill:2012rf}).

In general there are two types of symmetric space solutions: $AdS$-solutions and pp-wave type solutions \cite{Cahen-Wallach}.\footnote{It is easy to prove that there are no $dS$-solutions, e.g. \cite{FigueroaO'Farrill:2012rf}.} Since the latter are very simple and typically arise as Penrose limits of the former we will concentrate on the $AdS$-solutions. These are also of great interest from the point of view of holography \cite{Maldacena:1997re}.\footnote{However the new examples we find are non-supersymmetric and a recent conjecture \cite{Ooguri:2016pdq,Freivogel:2016qwc} suggests that they should therefore be unstable in string theory.} In particular these backgrounds are interesting since symmetric spaces are often associated with integrable string sigma models. In the absence of NSNS flux the bosonic string action is simply a symmetric space sigma-model which is a well known type of integrable model.

The geometry of the solutions consist of an $AdS_n$ factor (recall that we restrict here to $n>2$) and a Riemannian factor. The Riemannian factor is a product of irreducible Riemannian factors. Up to dimension seven these are listed in table \ref{tab:riemannian} together with the invariant forms they allow. In addition there can of course be flat directions which we will denote as compact, e.g. $T^n$, although our analysis is local and does not distinguish compact and non-compact flat directions. Once we have found all $AdS_n$ solutions for $n>2$ we also analyze the conditions for supersymmetry and find as expected that the only supersymmetric solutions are ones that are already well known.

\begin{table}[ht]
\begin{center}
\begin{tabular}{r|l|l}
Dim. & Name & Invariant forms\\
\hline 2 & $S^2$ & 0,2\\
3 & $S^3$ & 0,3\\
4 & $\mathbbm{CP}^2$ & 0,2,4\\
4 & $S^4$ & 0,4\\
5 & $SLAG_3$ & 0,5\\
5 & $S^5$ & 0,5\\
6 & $\mathbbm{CP}^3$ & 0,2,4,6\\
6 & $G_{\mathbbm R}^+(2,5)$ & 0,2,4,6\\
6 & $S^6$ & 0,6\\
7 & $S^7$ & 0,7
\end{tabular}
\caption{Irreducible Riemannian symmetric spaces up to dimension seven. For each entry there is also a non-compact version. $SLAG_3=SU(3)/SO(3)$ is the Grassmannian of special Lagrangian planes in $\mathbbm{C}^3$ and $G_{\mathbbm R}^+(2,5)=Sp(2)/U(2)$ is the Grassmannian of oriented $2$-planes in $\mathbbm{R}^5$.}
\label{tab:riemannian}
\end{center}
\end{table}

\begin{table}[ht]
\begin{center}
\begin{tabular}{rlcccc}
 &  & \multicolumn{2}{c}{Flux moduli} & &\\
 & Solution & IIA & IIB & SUSY & Comments\\
\hline
1. & $AdS_5\times S^5$ & $\times$ & - & 32 & \\
2. & $AdS_5\times SLAG_3$ & $\times$ & - & - & \\
3. & $AdS_5\times\mathbbm{CP}^2\times S^1$ & - & $\times$ & - & T-dual to 1\\
4. & $AdS_5\times S^3\times S^2$ & $\times$ & - & - & T-dual to special case of 5\\
5. & $AdS_5\times S^2\times S^2\times S^1$ & 1 & $\times$ & - & \\
\end{tabular}
\caption{The 5 symmetric $AdS_5$ solutions. The number of free parameters entering the fluxes are given for the type IIA and type IIB solution respectively with '-' denoting no free parameters and '$\times$' denoting that no solution exists. The number of supersymmetries preserved by the solution is also given.}
\label{tab:AdS5}
\end{center}
\end{table}
\begin{table}[ht]
\begin{center}
\begin{tabular}{rlcccc}
 &  & \multicolumn{2}{c}{Flux moduli} & &\\
 & Solution & IIA & IIB & SUSY & Comments\\
\hline
6. & $AdS_4\times\mathbbm{CP}^3$ & 1(-) & $\times$ & -(24) & (massive) IIA\\
7. & $AdS_4\times G_{\mathbbm R}^+(2,5)$ & 1 & $\times$ & - & (massive) IIA\\
8. & $AdS_4\times S^4\times S^2$ & 1 & $\times$ & - & (massive) IIA\\
9. & $AdS_4\times\mathbbm{CP}^2\times S^2$ & 2 & $\times$ & - & (massive) IIA\\
10. & $AdS_4\times S^3\times S^3$ & - & $\times$ & - & massive IIA\\
11. & $AdS_4\times S^3\times S^2\times S^1$ & $\times$ & - & - & T-dual to 10 and 13\\
12. & $AdS_4\times S^2\times S^2\times S^2$ & 3 & $\times$ & - & (massive) IIA\\
13. & $AdS_4\times S^2\times S^2\times T^2$ & - & $\times$ & - &\\
\end{tabular}
\caption{The 8 symmetric $AdS_4$ solutions. The number of free parameters entering the fluxes are given for the type IIA and type IIB solution respectively with '-' denoting no free parameters and '$\times$' denoting that no solution exists. The number of supersymmetries preserved by the solution is also given. The notation in the first line means that there is a subbranch of the solution with no free parameters which is supersymmetric. The comment '(massive) IIA' means that the general solution has non-zero Romans mass but there is a subbranch where it vanishes.}
\label{tab:AdS4}
\end{center}
\end{table}
\begin{table}[ht]
\begin{center}
\begin{tabular}{rlcccc}
 &  & \multicolumn{2}{c}{Flux moduli} & &\\
 & Solution & IIA & IIB & SUSY & Comments\\
\hline
14. & $AdS_3\times S^5\times H^2$ & $\times$ & - & - & \\
15. & $AdS_3\times SLAG_3\times H^2$ & $\times$ & - & - & \\
16. & $AdS_3\times\mathbbm{CP}^2\times S^2(H^2)\times S^1$ & 1 & $\times$ & - & \\
17. & $AdS_3\times\mathbbm{CP}^2\times T^3$ & - & $\times$ & - & \\
18. & $AdS_3\times S^3\times S^3\times S^1$ & 2 & 2 & 16 & T-dual\\
19. & $AdS_3\times S^3\times S^2\times H^2$ & $\times$ & 1 & - & \\
20. & $AdS_3\times S^3\times S^2\times T^2$ & 1 & 1 & 8 & T-dual\\
21. & $AdS_3\times S^3\times T^4$ & 1 & 2 & 16 & T-dual\\
22. & $AdS_3\times S^2\times S^2\times S^2(H^2)\times S^1$ & 3 & $\times$ & - & \\
23. & $AdS_3\times S^2\times S^2\times T^3$ & 2/2/2/3(1) & $\times$/$\times$/1/2(1) & -/-/-/-(8) & \\
24. & $AdS_3\times S^2\times T^5$ & 1 & - & 8 & T-dual
\end{tabular}
\caption{The 11 symmetric $AdS_3$ solutions. The number of free parameters entering the fluxes are given for the type IIA and type IIB solution respectively with '-' denoting no free parameters and '$\times$' denoting that no solution exists. The number of supersymmetries preserved by the solution is also given. {The comment ''T-dual'' means that the type IIA and IIB solution are related by T-duality on the torus (note that the number of free parameters in the flux can change if it becomes possible, after T-duality, to remove a parameter (locally) by a rotation of the torus directions).} The notation $S^2(H^2)$ means that either the compact $S^2$ or non-compact version $H^2$ can occur depending on the values of the parameters. Solution 23 has four distinct branches separated by a '/', the three-parameter branch has a one-parameter subbranch which preserves 8 supersymmetries.}
\label{tab:AdS3}
\end{center}
\end{table}

The result of our analysis is summarized in tables \ref{tab:AdS5}--\ref{tab:AdS3}. The solutions are numbered 1--24 and since two of the solutions allow also for a version with a non-compact Riemannian part there are 26 distinct geometries which break down as follows
\begin{center}
\begin{tabular}{cccc}
& IIA & IIB & Total\\
\hline
$AdS_5$ &  2 & 3 & 5\\
$AdS_4$ &  7 & 1 & 8\\
$AdS_3$ & 10 & 8 & 13
\end{tabular}
\end{center}
In many cases the type IIA and type IIB solutions are T-dual to each other. Many of the (non-massive) type IIA solutions arise by dimensional reduction of eleven-dimensional symmetric space solutions in \cite{Wulff:2016vqy}. {Except for solutions 23 and 24 in table \ref{tab:AdS3} (in their most general form) all the other type IIB solutions were already obtained in \cite{FigueroaO'Farrill:2012rf}. Their supersymmetries however were not previously analyzed.} Also most, if not all, of the non-massive type IIA $AdS_4$ solutions were found in the early 1980's due to the interest in compactifications down to four dimensions \cite{Watamura:1983ht,Volkov:1984yw,Sorokin:1985np,Campbell:1984zc,Giani:1984wc} (see also the review \cite{Duff:1986hr} although it deals mostly with $D=11$ supergravity). {Some of the non-supersymmetric solutions have certainly appeared in the literature before, e.g. $AdS_5\times\mathbbm{CP}^2\times S^1$ \cite{Duff:1998us}, but to my knowledge they have not been analyzed systematically apart from the previously mentioned works. The situation is of course different with the supersymmetric solutions.}

There are 7 solutions which admit supersymmetries. They are well known and arise as near horizon geometries of $D3$-branes ($AdS_5\times S^5$) \cite{Schwarz:1983qr,Gibbons:1993sv}, dimensional reduction of the near horizon of $M2$-branes ($AdS_4\times\mathbbm{CP}^3$) \cite{Nilsson:1984bj,Gibbons:1993sv} or as near horizon geometries of various intersecting brane configurations \cite{Papadopoulos:1996uq,Tseytlin:1996bh,Gauntlett:1996pb,Tseytlin:1996hi,Cowdall:1998bu,Boonstra:1998yu}. Furthermore these backgrounds have the nice property that the superstring is (at least classically) integrable \cite{Bena:2003wd,Arutyunov:2008if,Stefanski:2008ik,Sorokin:2010wn,Cagnazzo:2011at,Babichenko:2009dk,Sundin:2012gc,Cagnazzo:2012se,Sundin:2013uca,Wulff:2014kja,Wulff:2015mwa}.

The outline of the paper is as follows. In section \ref{sec:solutions} we give the supergravity equations that we need to solve in the case of symmetric spaces and then proceed to solve them for $AdS_n$ backgrounds starting with the highest $n$ and working out the type IIA case first and then the type IIB case. In section \ref{sec:susy} we analyze the supersymmetry of the solutions and we end with some conclusions. In the appendix we work out the Riemann tensor for some symmetric spaces.

\section{Type II symmetric space solutions}\label{sec:solutions}
As already remarked, when one restricts to symmetric space solutions the supergravity equations become algebraic since all terms with derivatives of field strengths or the dilaton vanish. We will use the supergravity conventions of \cite{Wulff:2013kga}. The type II supergravity equations of motion in these conventions can be found for example in Appendix A of \cite{Borsato:2016zcf}. In the case of symmetric spaces they reduce to
\begin{align}
&\mbox{\bf IIA}:\nonumber\\
&\quad R_{ab}=
\tfrac12\langle i_aH,i_bH\rangle
+\tfrac12\langle i_aF^{(2)},i_bF^{(2)}\rangle
+\tfrac12\langle i_aF^{(4)},i_bF^{(4)}\rangle
-\tfrac14((F^{(0)})^2+|F^{(2)}|^2+|F^{(4)}|^2)\eta_{ab}\,,
%\label{eq:Einstein}
\nonumber\\
\nonumber\\
&\quad 5(F^{(0)})^2+3|F^{(2)}|^2+|F^{(4)}|^2=2|H|^2\,,\quad
F^{(0)}*F^{(2)}-F^{(2)}\wedge*F^{(4)}+\tfrac12F^{(4)}\wedge F^{(4)}=0\,,
%2(i_{F^{(2)}}F^{(4)})_{ab}=F^{(2)}_{cd}F^{(4)}_{abcd}
%2*i_{F^{(2)}}F^{(4)}=-2F^{(2)}\wedge *F^{(4)}
\label{eq:IIA}\\
\nonumber\\
&\quad H\wedge *F^{(4)}=0\,,\quad H\wedge F^{(n)}=0\quad (n=0,2,4)\,,\nonumber\\
%\nonumber\\
\end{align}
\begin{align}
&\mbox{\bf IIB}:\nonumber\\
&\quad R_{ab}=
\tfrac12\langle i_aH,i_bH\rangle
+\tfrac12F^{(1)}_aF^{(1)}_b
+\tfrac12\langle i_aF^{(3)},i_bF^{(3)}\rangle
+\tfrac14\langle i_aF^{(5)},i_bF^{(5)}\rangle
-\tfrac14(|F^{(1)}|^2+|F^{(3)}|^2)\eta_{ab}\,,
\nonumber\\
\nonumber\\
&\quad 2|F^{(1)}|^2+|F^{(3)}|^2=|H|^2\,,\quad F^{(1)}\wedge*F^{(3)}+F^{(5)}\wedge F^{(3)}=0\,,\quad F^{(5)}=*F^{(5)}\,,
\label{eq:IIB}\\
\nonumber\\
&\quad H\wedge *F^{(3)}=0\,,\quad H\wedge F^{(n)}=0\quad(n=1,3,5)\,.
\nonumber
\end{align}
We will refer to the first line as the Einstein equation, the second line as the $FF$-equations and the third line as the $HF$-equations. The Ricci tensor is defined as $R_{ab}=R_{ac}{}^c{}_b$ and for each irreducible factor is just const.$\times\eta_{ab}(\delta_{ab})$, the NSNS field strength is $H=dB$, while the RR field strengths $F^{(n)}$ are defined in the usual way in terms of RR potentials $C^{(n-1)}$ with $F^{(0)}$ the Romans mass parameter. Since the dilaton is simply a constant we have absorbed it into the RR fields. The inner product on forms is defined as
\begin{equation}
\langle\alpha,\beta\rangle=\frac{1}{n!}\alpha_{a_1\cdots a_n}\beta^{a_1\cdots a_n}\,,\qquad |\alpha|^2=\langle\alpha,\alpha\rangle\,,
\end{equation}
where the components are defined as $\alpha=\frac{1}{n!}e^{a_n}\wedge\cdots\wedge e^{a_1}\alpha_{a_1\cdots a_n}$. We also define the contraction
\begin{equation}
i_a\alpha=i_{\partial_a}\alpha=\frac{1}{(n-1)!}e^{a_n}\wedge\cdots\wedge e^{a_2}\alpha_{aa_2\cdots a_n}\,.
\end{equation}
The Hodge dual is defined as
\begin{equation}
*\alpha=\frac{1}{(10-n)!n!}e^{a_{10-n}}\wedge\cdots\wedge e^{a_1}\varepsilon_{a_1...a_{10-n}}{}^{b_1\cdots b_n}\alpha_{b_n\cdots b_1}
\,,\qquad
**\alpha=\alpha\,,
\end{equation}
where $\varepsilon^{0123456789}=1$.

Let us make a few observations about these equations. The first of the $FF$-equations in (\ref{eq:IIA}) and (\ref{eq:IIB}) imply that if $H=0$ we must have RR-flux on $AdS$ since otherwise all fluxes vanish and no $AdS$ solution is possible. For type IIA this implies that $AdS_n$ solutions with $n>4$ must have non-zero $H$ and therefore no Romans mass $F^{(0)}=0$. In particular such geometries must have an invariant three-form. For type IIB the same is true except for in the $AdS_5$ case since in that case RR flux on $AdS_5$ is possible.

Furthermore it is not hard to prove (e.g. \cite{FigueroaO'Farrill:2012rf}) that for a solution of the form $AdS_n\times M_{10-n}$ the Riemannian part $M_{10-n}$ must have positive scalar curvature, in particular it cannot be flat.

Another useful observation is that solutions with a $\mathbbm{CP}^2$ factor can be thought of as special cases of solutions with an $S^2\times S^2$ factor. This is because the supergravity equations are essentially only sensitive to the invariant forms and the invariant forms of the former can be thought of as generated by a linear combination of those of the latter. Using this fact saves us having to analyze cases with $\mathbbm{CP}^2$ factors separately. Note that the same is true also for $\mathbbm{CP}^3$ and $G_{\mathbbm R}^+(2,5)$ solutions which can be treated as special cases of $S^2\times S^2\times S^2$ solutions but in this case there are so few examples that this observation is not very helpful. For similar reasons solutions with an $S^5$ or $SLAG_3$ factor can be thought of as special cases of solutions with an $S^3\times S^2$ factor.

Since we are solving quadratic equations for the parameters appearing in the ansatz for the fluxes we typically get various $\pm$ signs appearing in the solution. Often these are unimportant and can be removed. In particular the overall sign of $H$ or the overall sign of all RR fluxes is unimportant since it does not affect the equations of motion or supersymmetry equations, see section \ref{sec:susy}. The same is true for any change of sign of the invariant forms that \emph{preserves} the overall orientation, e.g. flipping the sign of two flat coordinates (but not one). The reason the orientation must be preserved is the appearance of the $\varepsilon$-symbol through the Hodge dual. We will make use of these facts to remove the inconsequential signs in the solutions.

All solutions are given in equations with the geometry in boldface together with the corresponding fluxes. The conditions on the parameters appearing in the fluxes are given in the equation or just below, e.g. (\ref{eq:5}). We also give the curvatures of the different irreducible factors as the number multiplying $\eta_{ab}(\delta_{ab})$ in the Ricci tensor. We use the following notation for the invariant forms appearing
\begin{itemize}
	\item $\nu$ denotes the unit volume form on $AdS$ with $|\nu|^2=-1$
	\item $\sigma$ denotes the unit volume form on spheres with $|\sigma|^2=1$
	\item $\omega$ denotes the K\"ahler form on $\mathbbm{CP}^n$ or $G_{\mathbbm R}^+(2,5)$ satisfying $|\omega|^2=n$ and which we take to have the form $\omega=e^2\wedge e^1+e^4\wedge e^3+\ldots$
	\item $\vartheta^{12\cdots n}$ denotes the basic forms on flat directions, i.e. $\vartheta^{12\cdots n}=dx^1\wedge dx^2\wedge\ldots\wedge dx^n$
\end{itemize}

Since we have already argued that $AdS_n$ solutions with $n>5$ must have an invariant three-form the highest possible dimension is $n=7$. We will analyze the type IIA case first and then the type IIB case.

\iffalse
\section{Symmetric space solutions of eleven-dimensional supergravity}
The equations of motion of eleven-dimensional supergravity take the form
\begin{equation}
R_{ab}=\tfrac12\langle i_aF,i_bF\rangle-\tfrac16|F|^2\eta_{ab}\,,\qquad d*F=-\tfrac12F\wedge F\,.
\label{eq:Einstein}
\end{equation}
where $F=dC$, with $C$ the three-form potential of eleven-dimensional supergravity.

Specializing to symmetric space solutions $F$ must be constant, $d*F=0$, so that the corresponding equation of motion reduces to the algebraic equation $F\wedge F=0$. The Einstein equation also becomes algebraic by noting that for an irreducible symmetric space any symmetric bilinear form, e.g. the Ricci tensor, is proportional to the metric. The proportionality constant is then determined in terms of $F$ by the Einstein equation.
\fi

\subsection{$AdS_7$ solutions: IIA}
For $AdS_7\times S^3$ the non-vanishing fluxes are $H=f_1\sigma$ and $F^{(0)}=f_2$. Since $f_1$ cannot vanish we must have $f_2=0$ but using (\ref{eq:IIA}) this implies vanishing $AdS_7$ curvature. For $AdS_7\times S^2\times S^1$ the fluxes are $H=f_1\sigma\wedge\vartheta^1$, $F^{(2)}=f_2\sigma$ and $F^{(0)}=f_3$. The Einstein equation in the $S^1$ direction gives $f_1^2=\frac12f_3^2+\frac12f_2^2$ which is inconsistent with the first $FF$ equation which says that $5f_3^2+3f_2^2=2f_1^2$.

\subsection{$AdS_7$ solutions: IIB}
For $AdS_7\times S^3$ the non-vanishing fluxes are $H=f_1\sigma$ and $F^{(3)}=f_2\sigma$. The condition $H\wedge*F^{(3)}=0$ implies $f_2=0$ but then $H$ must also vanish giving a contradiction. For $AdS_7\times S^2\times S^1$ the fluxes are $H=f_1\sigma\wedge\vartheta^1$, $F^{(3)}=f_2\sigma\wedge\vartheta^1$ and $F^{(1)}=f_3\vartheta^1$. Again we find $f_2=0$ but now $f_1^2=2f_3^2$ from the $FF$-equations in (\ref{eq:IIB}). The Einstein equation in the flat direction then implies $f_1=f_3=0$ and there is no solution.

\subsection{$AdS_6$ solutions: IIA}
For $AdS_6\times S^3\times S^1$ with $H=f_1\sigma$ ($f_1\neq0$) and $F^{(4)}=f_2\sigma\wedge\vartheta^1$ the condition $H\wedge*F^{(4)}=0$ forces $f_2=0$ and (\ref{eq:IIA}) implies vanishing $AdS$ curvature. The only other possible geometry with an invariant three-form is $AdS_6\times S^2\times T^2$. The flux takes the form $H=f_1\sigma\wedge\vartheta^1$, $F^{(4)}=f_2\sigma\wedge\vartheta^{12}$, $F^{(2)}=f_3\sigma+f_4\vartheta^{12}$ and $F^{(0)}=0$. Since $f_1\neq0$ the condition $H\wedge*F^{(4)}=0$ forces $f_2=0$. The difference of the Einstein equations in the two flat directions then gives $f_1=0$, a contradiction.

\subsection{$AdS_6$ solutions: IIB}
For $AdS_6\times S^3\times S^1$ the flux is $H=f_1\sigma$, $F^{(3)}=f_2\sigma$ and $F^{(1)}=f_3\vartheta^1$ but the $HF$-equations in (\ref{eq:IIB}) force $f_2=f_3=0$ and again the $AdS$ curvature vanishes. For $AdS_6\times S^2\times T^2$ the flux is $H=f_1\sigma\wedge\vartheta^1$, $F^{(3)}=\sigma\wedge(f_2\vartheta^1+f_3\vartheta^2)$ and $F^{(1)}=f_4\vartheta^1+f_5\vartheta^2$. Since $f_1\neq0$ we find from the $HF$-equations $f_2=f_5=0$ while the $FF$-equations give $f_1^2=2f_4^2+f_3^2$. The difference of the Einstein equations in the flat directions gives $f_1^2+f_4^2-f_3^2=0$ so that $f_4=0$ but then the Einstein equation in the 2-direction gives $f_3=f_1=0$, a contradiction.

\subsection{$AdS_5$ solutions: IIA}
For $AdS_5\times S^3\times S^2$ with non-zero $H$ we find from (\ref{eq:IIA}) $F^{(0)}=F^{(2)}=0$ and no solution. For $AdS_5\times S^3\times T^2$ we can have also non-zero $F^{(4)}$. We must have $H\neq0$ which again forces $F^{(0)}=F^{(2)}=0$ and, from $H\wedge*F^{(4)}=0$, also $F^{(4)}=0$ and no solution.

This leaves backgrounds with $S^2$ (or $\mathbbm{CP}^2$) factors and flat directions only. For $AdS_5\times S^2\times S^2\times S^1$ the flux is $H=(f_1\sigma_1+f_2\sigma_2)\wedge\vartheta^1$,  with say $f_1\neq0$, $F^{(4)}=f_3\sigma_1\wedge\sigma_2$, $F^{(2)}=f_4\sigma_1+f_5\sigma_2$ and $F^{(0)}=0$. The condition $H\wedge F^{(2)}=0$ gives $f_1f_5+f_2f_4=0$. The Einstein equation in the flat direction gives
\begin{equation}
2f_1^2+2f_2^2=f_3^2+f_4^2+f_5^2
\end{equation}
and the first $FF$-equation then forces $f_4=f_5=0$. We find the solution
\begin{equation}
\bm{AdS_5\times S^2\times S^2\times S^1}\,,\qquad
H=(f_1\sigma_1+f_2\sigma_2)\wedge\vartheta^1\,,\quad
F^{(4)}=f_3\sigma_1\wedge\sigma_2\,,\quad
f_3^2=2f_1^2+2f_2^2\,.
\label{eq:5}
\end{equation}
The curvatures\footnote{Recall that we mean that $R_{ab}$ is this number times $\eta_{ab}(\delta_{ab})$ for the corresponding factor in the geometry.} are $-\frac12(f_1^2+f_2^2)$, $f_1^2+\frac12f_2^2$ and $f_2^2+\frac12f_1^2$ respectively. This lifts to a solution in eleven dimensions \cite{FigueroaO'Farrill:2011fj,Wulff:2016vqy}.

As a special case of this analysis we get the solution
\begin{equation}
\bm{AdS_5\times\mathbbm{CP}^2\times S^1}\,,\qquad
H=f\omega\wedge\vartheta^1\,,\quad
F^{(4)}=f\omega\wedge\omega\,.
\label{eq:3}
\end{equation}
The curvatures are $-f^2$ and $\frac32f^2$ respectively. This also lifts to a solution in eleven dimensions.

Finally we have $AdS_5\times S^2\times T^3$ with $H=f_1\sigma\wedge\vartheta^1+f_2\vartheta^{123}$, $F^{(4)}=\sigma\wedge\alpha$, $F^{(2)}=f_3\sigma+\beta$ and $F^{(0)}=0$ with $\alpha,\beta\in\Omega^2(T^3)$. The trace of the Einstein equation (\ref{eq:IIA}) in the torus directions gives
\begin{equation}
\tfrac12f_1^2
+\tfrac32f_2^2
+|\beta|^2
+|\alpha|^2
-\tfrac34(f_3^2+|\beta|^2+|\alpha|^2)=0\,,
\end{equation}
while the first $FF$-equation gives
\begin{equation}
3f_3^2+3|\beta|^2+|\alpha|^2=2(f_1^2+f_2^2)\,,
\end{equation}
which implies $\alpha=\beta=0$, $f_2=0$ and $f_1^2=\frac32f_3^2$. The Einstein equation in the 2-direction then gives $f_3=0$ and no solution.

\subsection{$AdS_5$ solutions: IIB}
For $AdS_5\times S^5$ the flux takes the form $F^{(5)}=f(\nu+\sigma)$ and the same for $S^5$ replaced by $SLAG_3$ and we find the solutions
\begin{equation}
\bm{AdS_5\times S^5}\,,\qquad\bm{AdS_5\times SLAG_3}\,,\qquad F^{(5)}=f(\nu+\sigma)\,.
\label{eq:1-2}
\end{equation}
The first is T-dual to the solution in (\ref{eq:3}) \cite{Duff:1998us}.

For $AdS_5\times S^4\times S^1$ the flux takes the form $F^{(5)}=f_1(\nu+\sigma\wedge\vartheta^1)$ and $F^{(1)}=f_2\vartheta^1$. The $FF$-equations in (\ref{eq:IIB}) force $f_2=0$ but then the Einstein equation in the flat direction forces also $f_1=0$. For $AdS_5\times S^3\times S^2$ the flux is $H=f_1\sigma_1$, $F^{(5)}=f_2(\nu+\sigma_1\wedge\sigma_2)$ and $F^{(3)}=f_3\sigma_1$. The $HF$-equations imply $f_1f_2=f_1f_3=0$ while the first $FF$-equation gives $f_1^2=f_3^2$ so that $f_1=f_3=0$ and we find the solution
\begin{equation}
\bm{AdS_5\times S^3\times S^2}\,,\qquad F^{(5)}=f_2(\nu+\sigma_1\wedge\sigma_2)\,,
\label{eq:4}
\end{equation}
where all factors have the same radius of curvature. This solution is related by T-duality on the Hopf fiber of $S^3$ to the special case of solution (\ref{eq:5}) with $f_2=0$.

For $AdS_5\times S^3\times T^2$ the flux is $H=f_1\sigma$, $F^{(5)}=f_2(\nu+\sigma\wedge\vartheta^{12})$, $F^{(3)}=f_3\sigma$ and $F^{(1)}=f_4\vartheta^1$. The $HF$-equations force $f_1f_2=f_1f_3=f_1f_4=0$ while the first $FF$-equation gives $f_1^2=2f_4^2+f_3^2$ so that $f_1=f_3=f_4=0$. The Einstein equation in the flat directions then forces also $f_2=0$.

For $AdS_5\times S^2\times S^2\times S^1$ the flux is $H=(f_1\sigma^1+f_2\sigma^2)\wedge\vartheta^1$, $F^{(5)}=f_3(\nu+\sigma_1\wedge\sigma_2\wedge\vartheta^1)$, $F^{(3)}=(f_4\sigma^1+f_5\sigma^2)\wedge\vartheta^1$ and $F^{(1)}=f_6\vartheta^1$. The $HF$-equations force $f_1f_3=f_2f_3=f_1f_4+f_2f_5=0$ while the $FF$-equations imply $f_1^2+f_2^2=2f_6^2+f_4^2+f_5^2$ and $f_3f_4+f_5f_6=f_3f_5+f_4f_6=0$. If $f_3\neq0$ all the others vanish and the Einstein equation in the flat direction gives a contradiction. Therefore $f_3=0$ and the Einstein equation in the flat direction gives $2f_1^2+2f_2^2+f_6^2+f_4^2+f_3^2=0$ and there is no solution.

It remains only to analyze $AdS_5\times S^2\times T^3$ for which the flux is $H=f_1\sigma\wedge\vartheta^1+f_2\vartheta^{123}$, $F^{(5)}=f_3(\nu-\sigma\wedge\vartheta^{123})$, $F^{(3)}=\sigma\wedge(f_4\vartheta^1+f_5\vartheta^2)+f_6\vartheta^{123}$ and $F^{(1)}=f_7\vartheta^1+f_8\vartheta^2+f_9\vartheta^3$. The first $FF$-equation says $f_1^2+f_2^2=2(f_7^2+f_8^2+f_9^2)+f_4^2+f_5^2+f_6^2$ while the trace of the Einstein equation in the flat directions gives
\begin{equation}
2f_1^2+6f_2^2+3f_3^2-f_4^2-f_5^2+3f_6^2-f_7^2-f_8^2-f_9^2=0
\end{equation}
and it is not hard to see that there is no solution.

\subsection{$AdS_4$ solutions: IIA}
Looking at (\ref{eq:IIA}) we can have $H=0$ provided that there is flux on $AdS_4$. With a six-dimensional irreducible Riemannian factor we have the possibilities $AdS_4\times S^6$, $AdS_4\times\mathbbm{CP}^3$ and $AdS_4\times G_{\mathbbm R}^+(2,5)$. The first can only have $F^{(4)}$ flux but this flux vanishes by the first $FF$-equation so this geometry is ruled out. For the other two the flux takes the form $H=0$, $F^{(4)}=f_1\nu+\frac12f_2\omega\wedge\omega$, $F^{(2)}=f_3\omega$ and $F^{(0)}=f_4$. The first $FF$-equation in (\ref{eq:IIA}) gives $f_1^2=5f_4^2+9f_3^2+3f_2^2$ while the second $FF$-equation gives $f_1f_2-2f_2f_3-f_3f_4=0$ leading to the solution
\begin{align}
&\bm{AdS_4\times\mathbbm{CP}^3}\,,\qquad\bm{AdS_4\times G_{\mathbbm R}^+(2,5)}\,,\qquad
F^{(4)}=f_1\nu+\tfrac12f_2\omega\wedge\omega\,,\quad
F^{(2)}=f_3\omega\,,\quad
F^{(0)}=f_4\,,
\nonumber\\
&\qquad
f_2=f_3\big(f_1^2-9f_3^2\big)^{1/2}\big(5(f_1-2f_3)^2+3f_3^2\big)^{-1/2}\,,
\nonumber\\
&\qquad
f_4=(f_1-2f_3)\big(f_1^2-9f_3^2\big)^{1/2}\big(5(f_1-2f_3)^2+3f_3^2\big)^{-1/2}\,.
%f_1^2>=9f_3^2
\label{eq:6-7}
\end{align}
The curvatures are $-\frac32(f_4^2+2f_3^2+f_2^2)$ and $f_2^2+2f_3^2+f_4^2$. In the special case $f_2=f_4=0$ and $f_1=\pm 3f_3$ this background lifts to $AdS_4\times S^7$ in eleven dimensions \cite{Nilsson:1984bj}.

With a five-dimensional irreducible Riemannian factor we can only have non-zero $F^{(4)}$ but again the first $FF$-equation forces the flux to vanish. The next possibility is $AdS_4\times S^4\times S^2$ with flux $H=0$, $F^{(4)}=f_1\nu+f_2\sigma_1$, $F^{(2)}=f_3\sigma_2$ and $F^{(0)}=f_4$. The first $FF$-equation gives $f_1^2=5f_4^2+3f_3^2+f_2^2$ and the second $FF$-equation gives $f_4f_3-f_1f_2=0$ and we find the solution
\begin{align}
&\bm{AdS_4\times S^4\times S^2}\,,\qquad
F^{(4)}=f_1\nu+f_2\sigma_1\,,\quad
F^{(2)}=f_3\sigma_2\,,\quad
F^{(0)}=f_4\,,
\nonumber\\
&\qquad f_2=f_4f_3/f_1\,,\quad
f_1^2=\tfrac12\Big(5f_4^2+3f_3^2+\sqrt{(5f_4^2+3f_3^2)^2+4f_4^2f_3^2}\,\Big)\,,
\label{eq:8}
\end{align}
with curvatures $-\frac12(f_2^2+2f_3^2+3f_4^2)$, $\frac12(f_2^2+f_3^2+2f_4^2)$ and $f_3^2+f_4^2$. Since the curvature of the $S^2$ cannot vanish there cannot be a solution with $S^2$ replaced by $T^2$. In the special case $f_2=f_4=0$, $f_1=\pm\sqrt3f_3$ this lifts to the $AdS_4\times S^4\times S^3$ solution in eleven dimensions \cite{FigueroaO'Farrill:2011fj,Wulff:2016vqy}.

The next possibility is $AdS_4\times S^3\times S^3$ with flux $H=f_1\sigma_1+f_2\sigma_2$, $F^{(4)}=f_3\nu$ and $F^{(0)}=f_4$. We have either $H=0$ or $f_3=f_4=0$ but the latter is inconsistent with the first $FF$-equation and we conclude that $f_1=f_2=0$ and we have the solution
\begin{equation}
\bm{AdS_4\times S^3\times S^3}\,,\qquad
F^{(4)}=\sqrt5f\nu\,,\quad
F^{(0)}=f\,.
\label{eq:10}
\end{equation}
The curvatures are $-\frac32f^2$, $f^2$ and $f^2$.

For $AdS_4\times S^3\times S^2\times S^1$ the flux takes the form $H=f_1\sigma_1+f_2\sigma_2\wedge\vartheta^1$, $F^{(4)}=f_3\nu+f_4\sigma_1\wedge\vartheta^1$, $F^{(2)}=f_5\sigma_2$ and $F^{(0)}=f_6$. If $H\neq0$ we find $f_1f_5=0$ and $f_6=f_3=0$. The Einstein equation in flat direction then gives $f_5^2=2f_2^2+f_4^2$ while the first $FF$-equation gives $3f_5^2+f_4^2=2(f_1^2+f_2^2)$ and there is no non-trivial solution. We conclude that $f_1=f_2=0$ and the Einstein equation in the flat direction gives $f_3^2+f_4^2=f_5^2+f_6^2$ while the first $FF$-equation gives $f_3^2=5f_6^2+3f_5^2+f_4^2$ and again there is no non-trivial solution. For $AdS_4\times S^3\times T^3$ the flux takes the form $H=f_1\sigma+f_2\vartheta^{123}$, $F^{(4)}=f_3\nu+f_4\sigma\wedge\vartheta^1$, $F^{(2)}=\alpha$ and $F^{(0)}=f_5$ with $\alpha\in\Omega^2(T^3)$. If $H\neq0$ we again have $f_3=f_5=0$ and $f_1\alpha=0$. The trace of the Einstein equation in the flat directions gives
\begin{equation}
f_4^2=6f_2^2+|\alpha|^2\,,
\end{equation}
while the first $FF$-equation gives
\begin{equation}
3|\alpha|^2+f_4^2=2f_1^2+2f_2^2\,,
\end{equation}
so that $f_1^2=2|\alpha|^2+2f_2^2$ and since $f_1$ cannot vanish we must have $\alpha=0$. The Einstein equation in the 2-direction then forces $f_4=0$ which is inconsistent. We conclude that $f_1=f_2=0$ and the trace of the Einstein equation in the flat directions gives $f_4^2+3f_5^2=|\alpha|^2+3f_3^2$ while the first $FF$-equation gives $f_3^2=5f_5^2+3|\alpha|^2+f_4^2$ and again there is no non-trivial solution.

It remains to analyze geometries with only $S^2$ factors and flat directions (recall that $\mathbbm{CP}^2$ solutions follow as a special case). With three $S^2$ factors the flux takes the form $F^{(4)}=f_0\nu+\frac12f_{ij}\sigma_i\wedge\sigma_j$, $F^{(2)}=f_i\sigma_i$ and $F^{(0)}=f_4$. The $FF$-equations give
\begin{align}
&f_0^2=5f_4^2+3(f_1^2+f_2^2+f_3^2)+f_{12}^2+f_{13}^2+f_{23}^2\,,\qquad
f_4f_1+f_2f_{12}+f_3f_{13}-f_0f_{23}=0\,,
\nonumber\\
&f_4f_2+f_1f_{12}+f_3f_{23}-f_0f_{13}=0\,,\qquad
f_4f_3+f_1f_{13}+f_2f_{23}-f_0f_{12}=0\,.
\end{align}
The solution is (rescaling $f_{1,2,3}\rightarrow f_0f_{1,2,3}$)
\begin{align}
&\bm{AdS_4\times S^2\times S^2\times S^2}\,,\qquad 
F^{(4)}=f_0\nu+\tfrac12f_{ij}\sigma_i\wedge\sigma_j\,,\quad
F^{(2)}=f_0f_i\sigma_i\,,\quad
F^{(0)}=f_4\,,
\nonumber\\
&\qquad
f_0^2(1-3[f_1^2+f_2^2+f_3^2])=5f_4^2+f_{12}^2+f_{13}^2+f_{23}^2\,,
\nonumber\\
&\qquad
f_{12}=f_4(2f_1f_2+f_3[1+f_1^2+f_2^2-f_3^2])/(1-f_1^2-f_2^2-f_3^2-2f_1f_2f_3)\,,
\nonumber\\
&\qquad
f_{13}=f_4(2f_1f_3+f_2[1+f_1^2-f_2^2+f_3^2])/(1-f_1^2-f_2^2-f_3^2-2f_1f_2f_3)\,,
\nonumber\\
&\qquad
f_{23}=f_4(2f_2f_3+f_1[1-f_1^2+f_2^2+f_3^2])/(1-f_1^2-f_2^2-f_3^2-2f_1f_2f_3)\,.
\label{eq:12}
\end{align}
The curvatures are $-\frac14(f_0^2(1+f_i^2)+f_4^2+f_{ij}^2)$, $\frac12(2f_4^2+f_{12}^2+f_{13}^2+f_0^2[2f_1^2+f_2^2+f_3^2])$, $\frac12(2f_4^2+f_{12}^2+f_{23}^2+f_0^2[f_1^2+2f_2^2+f_3^2])$ and $\frac12(2f_4^2+f_{13}^2+f_{23}^2+f_0^2[f_1^2+f_2^2+2f_3^2])$ respectively. In the special case $f_2=f_3=f_4=f_{ij}=0$ and $f_1=\frac{1}{\sqrt3}$ this lifts to $AdS_4\times S^3\times S^2\times S^2$ in $D=11$ \cite{FigueroaO'Farrill:2011fj,Wulff:2016vqy}.

The curvatures in the above solution are all strictly positive which means that there is no solution with one $S^2$ replaced by $T^2$ unless $H$ is non-zero. The flux is then as before but with $f_4=0$, $\sigma_3=\vartheta^{12}$ and $H=f_5\sigma_1\wedge\vartheta^1+\sigma_2\wedge(f_6\vartheta^1+f_7\vartheta^2)$. The $FF$ and $HF$-equations imply
\begin{align}
&f_1f_6+f_2f_5=f_1f_7=0\,,\qquad
f_5f_{13}+f_6f_{23}=0\,,\qquad
f_7f_{23}=0\,,\qquad
f_1f_{12}+f_3f_{23}-f_0f_{13}=0\,,
\nonumber\\
&f_1f_{13}+f_2f_{23}-f_0f_{12}=0\,,\qquad
f_2f_{12}+f_3f_{13}-f_0f_{23}=0\,,
\nonumber\\
&3(f_1^2+f_2^2+f_3^2)+f_{12}^2+f_{13}^2+f_{23}^2=f_0^2+2(f_5^2+f_6^2+f_7^2)\,,
\end{align}
while the Einstein equation in the flat directions gives
\begin{equation}
f_5^2+f_6^2-f_7^2=0\,,\qquad
f_0^2
-f_1^2
-f_2^2
+f_3^2
+2f_7^2
-f_{12}^2
+f_{13}^2
+f_{23}^2
=0\,,\qquad
f_6f_7=0\,.
\end{equation}
Clearly we must have $f_7\neq0$ since otherwise $H$ vanishes. This implies $f_1=f_6=f_{23}=0$ and we find the solution
\begin{align}
%f_0=f_1=f_2=f_6=f_{23}=f_{13}=0\,,\qquad
f_5^2=f_7^2=2f_3^2\,,\qquad
f_{12}^2=5f_3^2
\end{align}
and the rest vanishing, giving the solutions
\begin{equation}
\bm{AdS_4\times S^2\times S^2\times T^2}\,,\qquad
H=\sqrt2f_3(\sigma_1\wedge\vartheta^1\pm\sigma_2\wedge\vartheta^2)\,,\quad
F^{(4)}=\sqrt5f_3\sigma_1\wedge\sigma_2\,,\quad
F^{(2)}=f_3\vartheta^{12}\,.
\label{eq:13}
\end{equation}
The curvatures are respectively $-\frac32f_3^2$, $2f_3^2$ and $2f_3^2$.

As a special case of the above analysis we obtain the solution
\begin{align}
&\bm{AdS_4\times\mathbbm{CP}^2\times S^2}\,,\qquad
F^{(4)}=f_0\nu+\tfrac12f_{12}\omega^2+f_{13}\omega\wedge\sigma\,,\quad
F^{(2)}=f_0f_1\omega+f_0f_3\sigma\,,
\nonumber\\
&\qquad
F^{(0)}=f_4(1-f_3-2f_1^2)\,,\qquad
f_{12}=f_4(2f_1^2+f_3(1-f_3))\,,\quad
f_{13}=f_4f_1(1+f_3)\,,
\nonumber\\
&\qquad
f_0^2(1-3[2f_1^2+f_3^2])=f_4^2[5(1-f_3-2f_1^2)^2+(2f_1^2+f_3(1-f_3))^2+2f_1^2(1+f_3)^2]\,.
\label{eq:9}
\end{align}
The curvatures are $-\frac14(f_0^2(1+2f_1^2+f_3^2)+f_4^2+f_{12}^2+2f_{13}^2)$, $\frac12(2f_4^2+f_{12}^2+f_{13}^2+f_0^2[3f_1^2+f_3^2])$ and $f_4^2+f_{13}^2+f_0^2(f_1^2+f_3^2)$ respectively.

The only remaining possibility is $AdS_4\times S^2\times T^4$. The flux takes the form $H=\sigma\wedge\alpha+\beta$, $F^{(4)}=f_1\nu+\sigma\wedge\gamma+f_2\vartheta^{1234}$, $F^{(2)}=f_3\sigma+f_4\vartheta^{12}+f_5\vartheta^{34}$ and $F^{(0)}=f_6$ where $\alpha\in\Omega^1(T^4)$, $\beta\in\Omega^3(T^4)$ and $\gamma\in\Omega^2(T^4)$. The first $FF$-equation gives
\begin{equation}
5f_6^2+3f_3^2+3f_4^2+3f_5^2+|\gamma|^2+f_2^2=f_1^2+2|\alpha|^2+2|\beta|^2\,,
\end{equation}
while the trace of the Einstein equation in the flat directions gives
\begin{equation}
\tfrac12|\alpha|^2+\tfrac32|\beta|^2+f_2^2+f_1^2=f_3^2+f_6^2\,.
\end{equation}
These equations imply
\begin{equation}
|\alpha|^2=f_1^2+3f_2^2+f_3^2+3f_4^2+3f_5^2+3f_6^2+|\beta|^2+|\gamma|^2\,,
\end{equation}
so that $\alpha\neq0$ which implies $f_1=f_6=0$ but therefore also $f_3\neq0$ and from $H\wedge F^{(2)}=0$ we get $\beta=-(1/f_3)(f_4\vartheta^{12}+f_5\vartheta^{34})\wedge\alpha$. From the second $FF$-equation we find $\gamma=-(f_2/f_3)(f_4\vartheta^{34}+f_5\vartheta^{12})$ and $f_2f_4f_5=0$. We can now take, without loss of generality, $\alpha=f_7\vartheta^1+f_8\vartheta^3$ but the difference of the Einstein equation in the 1 and 2-direction forces $f_7=0$ and similarly we find $f_8=0$ but this contradicts the fact that $\alpha\neq0$ hence there is no solution.

\subsection{$AdS_4$ solutions: IIB}
With a six-dimensional irreducible Riemannian factor there is no way to put the flux. For $AdS_4\times S^5\times S^1$ the flux is $F^{(5)}=f_1(\nu\wedge\vartheta^1+\sigma)$ and $F^{(1)}=f_2\vartheta^1$. The first $FF$-equation in (\ref{eq:IIB}) forces $f_2=0$ and the Einstein equation in the flat direction forces $f_1=0$. The same applies with $S^5$ replaced by $SLAG_3$. For $AdS_4\times S^4\times S^2$ there is no way to put the flux. For $AdS_4\times S^4\times T^2$ the flux is $F^{(5)}=f_1(\nu\wedge\vartheta^1+\sigma\wedge\vartheta^2)$ and $F^{(1)}=f_2\vartheta^1$. Again we find $f_2=0$ and the Einstein equation forces $f_1=0$.

For $AdS_4\times S^3\times S^3$ the flux is $H=f_1\sigma_1+f_2\sigma_2$ and $F^{(3)}=f_3\sigma_1+f_4\sigma_2$. The $HF$-equations imply $f_1f_4-f_2f_3=f_1f_3+f_2f_4=0$ which makes either $H$ or $F^{(3)}$ vanish so that the first $FF$-equation forces the remaining flux to vanish. For $AdS_4\times S^3\times S^2\times S^1$ the flux is $H=f_1\sigma_1+f_2\sigma_2\wedge\vartheta^1$, $F^{(5)}=f_3(\nu\wedge\vartheta^1+\sigma_1\wedge\sigma_2)$, $F^{(3)}=f_4\sigma_1+f_5\sigma_2\wedge\vartheta^1$ and $F^{(1)}=f_6\vartheta^1$. The equations $H\wedge F^{(3)}=H\wedge*F^{(3)}=0$ imply $f_1f_5-f_2f_4=f_1f_4+f_2f_5=0$ which forces either $H=0$ of $F^{(3)}$=0 but the former implies the latter through the first $FF$-equation. Therefore $f_4=f_5=0$ and the remaining $HF$-equations imply $f_1f_6=f_1f_3=0$ while the $FF$-equations imply $f_1^2+f_2^2=2f_6^2$. We conclude that $f_1=0$ and the Einstein equation in the $S^1$ direction then says that $f_3^2=5f_6^2$ and we have the solution
\begin{align}
&\bm{AdS_4\times S^3\times S^2\times S^1}\,,\qquad
H=\sqrt2f\sigma_2\wedge\vartheta^1\,,\quad
F^{(5)}=\sqrt5f(\nu\wedge\vartheta^1+\sigma_1\wedge\sigma_2)\,,\quad
F^{(1)}=f\vartheta^1\,.
\label{eq:11}
\end{align}
The curvatures are $-\frac32f^2$, $f^2$ and $2f^2$ respectively. This solution is T-dual, via T-duality on the Hopf fiber of $S^3$, to both solution (\ref{eq:13}) and (\ref{eq:10}).

It remains only to analyze backgrounds with $S^2$ factors and flat directions. With three $S^2$ factors there is no way to put the flux. For $AdS_4\times S^2\times S^2\times T^2$ the flux takes the form $H=\sigma_1\wedge\alpha+\sigma_2\wedge\beta$, $F^{(5)}=f_1(\nu\wedge\vartheta^1+\sigma_1\wedge\sigma_2\wedge\vartheta^2)$, $F^{(3)}=\sigma_1\wedge\gamma+\sigma_2\wedge\delta$ and $F^{(1)}=\eta$ with $\alpha,\beta,\gamma,\delta,\eta\in\Omega^1(T^2)$. The trace of the Einstein equation in the flat directions gives $|\alpha|^2+|\beta|^2=0$ so that $H=0$ and the first $FF$-equation forces $F^{(1)}=F^{(3)}=0$ as well. Finally the Einstein equation in the flat directions forces $f_1=0$.

For $AdS_4\times S^2\times T^4$ the flux takes the form $H=\sigma\wedge\alpha+\beta$, $F^{(5)}=f_1(\nu\wedge\vartheta^1-\sigma\wedge\vartheta^{234})$, $F^{(3)}=\sigma\wedge\gamma+\delta$ and $F^{(1)}=\eta$ with $\alpha,\gamma,\eta\in\Omega^1(T^2)$ and $\beta,\delta\in\Omega^3(T^4)$. The first $FF$-equation says that
\begin{equation}
|\alpha|^2+|\beta|^2=2|\eta|^2+|\gamma|^2+|\delta|^2\,,
\end{equation}
while the trace of the Einstein equation in the flat directions gives
\begin{equation}
|\alpha|^2+3|\beta|^2+|\delta|^2+f_1^2=|\gamma|^2+|\eta|^2\,,
\end{equation}
forcing $f_1=0$ and $\beta=\delta=\eta=0$ but then the $HF$-equations force also $\alpha=\gamma=0$.

\subsection{$AdS_3$ solutions: IIA}
The only irreducible 7-dimensional factor is $S^7$ and the flux becomes $H=f_1\nu$, $F^{(0)}=f_2$ and the rest vanishing but that is incompatible with the first $FF$-equation in (\ref{eq:IIA}). For $S^6\times S^1$ the flux is $H=f_1\nu$, $F^{(4)}=f_2\nu\wedge\vartheta^1$, $F^{(0)}=f_3$ and the rest vanishing. The Einstein equation in the flat direction forces $f_2=f_3=0$ and there is no solution. For $\mathbbm{CP}^3\times S^1$ or $G_{\mathbbm R}^+(2,5)\times S^1$ the flux takes the form $H=f_1\nu+f_2\omega\wedge\vartheta^1$, $F^{(4)}=f_3\nu\wedge\vartheta^1+\frac12f_4\omega\wedge\omega$, $F^{(2)}=f_5\omega$ and $F^{(0)}=f_6$. The first $FF$-equation gives $5f_6^2+9f_5^2-f_3^2+3f_4^2=-2f_1^2+6f_2^2$ while the Einstein equation in the flat direction says that $6f_2^2-f_3^2-f_6^2-3f_4^2-3f_5^2=0$. This gives $6f_2^2=3f_6^2+6f_5^2+f_1^2+3f_4^2$ and $f_3^2=2f_6^2+3f_5^2+f_1^2$ so that $f_2$ cannot vanish which implies $f_6=0$. The condition $H\wedge F^{(4)}=0$ then forces $f_4=0$ and similarly $f_5=0$ but then $H\wedge*F^{(4)}=0$ gives $f_1f_3=0$ and there is no solution.

For $AdS_3\times S^5\times S^2$ the flux takes the form $H=f_1\nu$, $F^{(2)}=f_2\sigma$ and $F^{(0)}=f_3$ while $F^{(4)}$ vanishes. The $HF$ and $FF$-equations in (\ref{eq:IIA}) imply that $f_1f_2=f_1f_3=f_2f_3=0$ and $5f_3^2+3f_2^2=-2f_1^2$ which has no non-trivial solution. Replacing the $S^2$ by $T^2$ the flux takes the form $H=f_1\nu$, $F^{(4)}=f_2\nu\wedge\vartheta^1$, $F^{(2)}=f_3\vartheta^{12}$ and $F^{(0)}=f_4$. The $FF$ and $HF$-equations imply $f_1f_2=f_1f_3=f_1f_4=f_3f_4=0$ and $5f_4^2+3f_3^2-f_2^2=-2f_1^2$ giving $f_1=f_3f_4=0$ and $f_2^2=5f_4^2+3f_3^2$. The difference of the Einstein equations in the flat directions gives $f_2=0$ and no solution. The same applies with $S^5$ replaced by $SLAG_3$. The next possibility is $AdS_3\times S^4\times S^3$ with flux $H=f_1\nu+f_2\sigma_2$, $F^{(4)}=f_3\sigma_1$, $F^{(0)}=f_4$ and $F^{(2)}$ vanishing. The $HF$ and $FF$-equations imply $f_1f_4=f_2f_4=f_1f_3=f_2f_3=0$ and $5f_4^2+f_3^2=-2f_1^2+2f_2^2$ implying $f_3=f_4=0$ and $f_1=f_2$. However the Einstein equation then implies that the curvature of the $S^4$ vanishes giving a contradiction.

For $AdS_3\times S^4\times S^2\times S^1$ the flux is $H=f_1\nu+f_2\sigma_2\wedge\vartheta^1$, $F^{(4)}=f_3\nu\wedge\vartheta^1+f_4\sigma_1$, $F^{(2)}=f_5\sigma_2$ and $F^{(0)}=f_6$. The $HF$ and $FF$-equations imply $f_1f_6=f_2f_6=f_1f_5=f_1f_4=f_2f_4=f_1f_3=0$, $f_6f_5+f_3f_4=0$ and $5f_6^2+3f_5^2+f_4^2-f_3^2=-2f_1^2+2f_2^2$. The Einstein equation in the flat direction gives $2f_2^2-f_3^2-f_6^2-f_5^2-f_4^2=0$ and we find $f_1=f_4=f_6=0$, $f_3=f_2$ and $f_5=f_2$, but then the curvature of the $S^4$ vanishes giving a contradiction.

For $AdS_3\times S^4\times T^3$ the flux is, without loss of generality, $H=f_1\nu+f_2\vartheta^{123}$, $F^{(4)}=f_3\nu\wedge\vartheta^1+f_4\sigma_1$, $F^{(2)}=f_5\vartheta^{12}+f_6\vartheta^{23}$ and $F^{(0)}=f_7$. The $HF$ and $FF$-equations imply $f_1f_7=f_2f_7=f_1f_4=f_2f_4=f_1f_5=f_1f_6=f_1f_3=f_7f_5=0$, $f_7f_6+f_3f_4=0$ and $5f_7^2+3f_5^2+3f_6^2-f_3^2+f_4^2=-2f_1^2+2f_2^2$. The difference of the Einstein equations in the 2 and 3-directions gives $f_5=0$ which means that the flux has the same form as in the previous case and there is no solution.

For $AdS_3\times S^3\times S^3\times S^1$ the flux takes the form $H=f_1\nu+f_2\sigma_1+f_3\sigma_2$, $F^{(4)}=(f_4\nu+f_5\sigma_1+f_6\sigma_2)\wedge\vartheta^1$ and $F^{(0)}=f_7$ while $F^{(2)}$ vanishes. The $HF$ and $FF$-equations imply $f_1f_7=f_2f_7=f_3f_7=0$, $f_1f_5-f_4f_2=f_1f_6-f_4f_3=f_2f_6-f_5f_3=0$, $-f_1f_4+f_2f_5+f_3f_6=0$ and $5f_7^2-f_4^2+f_5^2+f_6^2=-2f_1^2+2f_2^2+2f_3^2$. The Einstein equation in the flat direction implies $-f_4^2+f_5^2+f_6^2-f_7^2=0$, which together with the other equations implies $f_7=0$ and we find the solution (rescaling $f_2$ and $f_3$ by $f_1$ and $f_5$ and $f_6$ by $f_4$)
\begin{equation}
\bm{AdS_3\times S^3\times S^3\times S^1}\,,\qquad
H=f_1(\nu+f_2\sigma_1+f_3\sigma_2)\,,\quad
F^{(4)}=f_4(\nu+f_2\sigma_1+f_3\sigma_2)\wedge\vartheta^1\,,
\label{eq:18}
\end{equation}
with $f_2^2+f_3^2=1$. The curvatures are $-\frac12(f_1^2+f_4^2)$, $\frac12(f_1^2+f_4^2)f_2^2$ and $\frac12(f_1^2+f_4^2)f_3^2$ respectively. This well-known solution clearly lifts to $D=11$.

For $AdS_3\times S^3\times S^2\times S^2$ the flux takes the form $H=f_1\nu+f_2\sigma_1$, $F^{(4)}=f_3\sigma_2\wedge\sigma_3$, $F^{(2)}=f_4\sigma_2+f_5\sigma_3$ and $F^{(0)}=f_6$. Since there is no RR flux on $AdS$ the first $FF$ equation implies $H\neq0$ so that $f_6=0$. The remaining $HF$ and $FF$-equations then give $f_3=f_4=f_5=0$ and $f_1=f_2$ but then the Einstein equation implies that the curvature of the $S^2$'s vanish giving a contradiction.

For $AdS_3\times S^3\times S^2\times T^2$ the flux takes the form $H=f_1\nu+f_2\sigma_1+\sigma_2\wedge(f_3\vartheta^1+f_4\vartheta^2)$, $F^{(4)}=f_5\nu\wedge\vartheta^1+\sigma_1\wedge(f_6\vartheta^1+f_7\vartheta^2)+f_8\sigma_2\wedge\vartheta^{12}$, $F^{(2)}=f_9\sigma_2+f_{10}\vartheta^{12}$ and $F^{(0)}=f_{11}$. Assume first that $f_2\neq0$. Then the $HF$-equations imply $f_9=f_{10}=f_{11}=0$ while the trace of the Einstein equation in the flat directions gives
\begin{equation}
\tfrac12f_3^2+\tfrac12f_4^2+\tfrac12f_8^2=0\,,
\end{equation}
so that $f_3=f_4=f_8=0$. The second $FF$-equation gives $f_5f_7=0$. If $f_5=0$ we can make a rotation in the (12)-plane to set $f_7=0$, so we conclude that $f_7=0$. The Einstein equation in the flat directions then implies $f_6^2=f_5^2$ but then the curvature of the $S^2$ vanishes giving a contradiction. 

It remains to analyze the case $f_2=0$. The first $FF$-equation gives
\begin{equation}
5f_{11}^2+3f_9^2+3f_{10}^2+f_8^2+f_6^2+f_7^2-f_5^2=-2f_1^2+2f_3^2+2f_4^2\,,
\end{equation}
while the Einstein equation in the flat directions gives
\begin{equation}
f_3^2+f_4^2+f_8^2-f_9^2+f_{10}^2-f_{11}^2=0\,,\qquad
f_3^2-f_4^2-f_5^2+f_6^2-f_7^2=0\,.
\end{equation}
These equations imply $f_3^2=2f_{11}^2+2f_{10}^2+f_9^2+f_8^2+f_7^2+f_1^2$. If $f_3=0$ the curvature of the $S^2$ vanishes so we must have $f_3\neq0$. Then the $HF$-equations imply $f_{11}=f_8=0$ and $f_5f_4=f_1f_6=f_1f_7=f_1f_5=f_5f_7=0$. Since from the above equations $f_5^2=5f_{10}^2+f_9^2+f_7^2+f_6^2+2f_1^2$ we must have $f_5\neq0$ and therefore $f_1=f_4=f_7=0$. Using this in the above equations we finally get $f_{10}=0$, $f_9^2=f_3^2$ and $f_5^2=f_3^2+f_6^2$ and the solution
\begin{equation}
\bm{AdS_3\times S^3\times S^2\times T^2}\,,\qquad
H=f_3\sigma_2\wedge\vartheta^1\,,\quad
F^{(4)}=(f_5\nu+f_6\sigma_1)\wedge\vartheta^1\,,\quad
F^{(2)}=f_3\sigma_2\,,
\label{eq:20}
\end{equation}
with $f_5^2=f_3^2+f_6^2$. The curvatures are $-\frac12f_5^2$, $\frac12f_6^2$ and $f_3^2$ respectively. This solution lifts to the $AdS_3\times S^3\times S^3\times T^2$ solution in $D=11$.

For $AdS_3\times S^3\times T^4$ the flux takes the form $H=f_1\nu+f_2\sigma+\alpha$, $F^{(4)}=f_3\nu\wedge\vartheta^1+\sigma\wedge\beta+f_4\vartheta^{1234}$, $F^{(2)}=\gamma$ and $F^{(0)}=f_5$ with $\alpha\in\Omega^3(T^4)$, $\beta\in\Omega^1(T^4)$ and $\gamma\in\Omega^2(T^4)$. Taking the trace of the Einstein equation in the flat directions gives
\begin{equation}
\tfrac32|\alpha|^2+\tfrac12f_3^2-\tfrac12|\beta|^2+f_4^2-f_5^2=0\,,
\end{equation}
while the first $FF$-equation gives
\begin{equation}
5f_5^2+3|\gamma|^2-f_3^2+|\beta|^2+f_4^2=-2f_1^2+2f_2^2+2|\alpha|^2\,.
\end{equation}
It is easy to see from these equations that if $f_2=0$ then $f_1=f_4=f_5=\alpha=\gamma=0$ and $f_3^2=|\beta|^2$. If on the other hand $f_2\neq0$ then the $HF$-equations imply that again $f_5=\gamma=0$ and from the above equations $|\beta|^2=3|\alpha|^2+f_3^2+2f_4^2$ and $2f_2^2=|\alpha|^2+3f_4^2+2f_1^2$ so that the previous case can be thought of a special case of this. The second $FF$-equation implies either $f_3=0$ or $\beta=f_6\vartheta^1$ for some $f_6$. However if $f_3=0$ we can still set $\beta=f_6\vartheta^1$ by a rotation. Therefore $\beta=f_6\vartheta^1$ and taking $\alpha=f_7\vartheta^{234}+f_8\vartheta^{123}$ the difference of the Einstein equation in the 1 and 2-directions now gives $f_6^2=f_3^2+f_7^2$ which together with the previous equations implies $f_4=f_7=f_8=0$ and we have the solution \begin{equation}
\bm{AdS_3\times S^3\times T^4}\,,\qquad
H=f_1(\nu+\sigma)\,,\quad
F^{(4)}=f_3(\nu+\sigma)\wedge\vartheta^1\,.
\label{eq:21}
\end{equation}
The curvatures are $-\frac12(f_1^2+f_3^2)$, $\frac12(f_1^2+f_3^2)$. This is again a well-known solution which also lifts to $D=11$.

It remains only to analyze geometries with $S^2$ factors and flat directions ($\mathbbm{CP}^2$ solutions arise as a special case as already noted). For $AdS_3\times S^2\times S^2\times S^2\times S^1$ the flux takes the form $H=f_1\nu+(f_2\sigma_1+f_3\sigma_2+f_4\sigma_3)\wedge\vartheta^1$, $F^{(4)}=f_5\nu\wedge\vartheta^1+f_6\sigma_1\wedge\sigma_2+f_7\sigma_1\wedge\sigma_3+f_8\sigma_2\wedge\sigma_3$, $F^{(2)}=f_9\sigma_1+f_{10}\sigma_2+f_{11}\sigma_3$ and $F^{(0)}=f_{12}$. The Einstein equation in the flat direction says that $|F|^2=2(|H|^2+f_1^2-f_5^2)$, where $|F|^2$ is the sum of the squares of the RR fluxes, while the first $FF$-equation says $|F|^2+2f_9^2+2f_{10}^2+2f_{11}^2+4f_{12}^2=2|H|^2$ so that $f_5^2=f_1^2+f_9^2+f_{10}^2+f_{11}^2+2f_{12}^2$. From the equation $H\wedge*F^{(4)}=0$ we get $f_1f_5=0$ and by the previous equation we have $f_1=0$. If $H=0$ the equation $|F|^2+2f_5^2=2|H|^2$ forces all flux to vanish so we must have $H\neq0$ and therefore $f_{12}=0$. The remaining $HF$ and $FF$-equations become
\begin{align}
&f_2f_8+f_3f_7+f_4f_6=0\,,\qquad
f_2f_{10}+f_3f_9=0\,,\qquad
f_2f_{11}+f_4f_9=0\,,\qquad
f_3f_{11}+f_4f_{10}=0\,,\qquad
\nonumber\\
&f_{10}f_6+f_{11}f_7-f_5f_8=0\,,\qquad
f_9f_6+f_{11}f_8-f_5f_7=0\,,\qquad
f_9f_7+f_{10}f_8-f_5f_6=0\,.
\end{align}
Assume $f_9\neq0$. In this case the above equations have a solution but one finds that the curvature of one of the $S^2$'s vanishes giving a contradiction. Therefore we must take $f_9=f_{10}=f_{11}=0$ and we get the solution
\begin{align}
&\bm{AdS_3\times S^2\times S^2\times S^2(H^2)\times S^1}\,,\qquad
H=(f_2\sigma_1+f_3\sigma_2+f_4\sigma_3)\wedge\vartheta^1\,,
\nonumber\\
&\qquad
F^{(4)}=f_6\sigma_1\wedge\sigma_2+f_7\sigma_1\wedge\sigma_3+f_8\sigma_2\wedge\sigma_3\,,\qquad
f_2f_8+f_3f_7+f_4f_6=0\,,
\nonumber\\
&\qquad
f_6^2+f_7^2+f_8^2=2(f_2^2+f_3^2+f_4^2)\,.
\label{eq:22}
\end{align}
The curvatures are $-\frac14(f_6^2+f_7^2+f_8^2)$, $\frac14(2f_2^2+f_6^2+f_7^2-f_8^2)$, $\frac14(2f_3^2+f_6^2+f_8^2-f_7^2)$ and $\frac14(2f_4^2+f_7^2+f_8^2-f_6^2)$ respectively and the sum of any two of the last three is positive. This solution also lifts to $D=11$ \cite{FigueroaO'Farrill:2011fj,Wulff:2016vqy}.

A special case of this analysis gives the solution
\begin{align}
&\bm{AdS_3\times\mathbbm{CP}^2\times S^2(H^2)\times S^1}\,,\qquad
H=(f_2\sigma+f_3\omega)\wedge\vartheta^1\,,\quad
F^{(4)}=(f_2\sigma-f_3\omega)\wedge\omega\,.
\label{eq:16}
\end{align}
The curvatures are $-\frac12(f_2^2+2f_3^2)$, $\frac32f_3^2$ and $f_2^2-f_3^2$. This solution again lifts to $D=11$ \cite{FigueroaO'Farrill:2011fj,Wulff:2016vqy}.

For $AdS_3\times S^2\times S^2\times T^3$ the flux takes the form $H=f_1\nu+\sigma_1\wedge\alpha+\sigma_2\wedge\beta+f_2\vartheta^{123}$, $F^{(4)}=f_3\nu\wedge\vartheta^1+f_4\sigma_1\wedge\sigma_2+\sigma_1\wedge\gamma+\sigma_2\wedge\delta$, $F^{(2)}=f_5\sigma_1+f_6\sigma_2+\zeta$ and $F^{(0)}=f_7$ with $\alpha,\beta\in\Omega^1(T^3)$ and $\gamma,\delta,\zeta\in\Omega^2(T^3)$. Taking the trace of the Einstein equation in the flat directions gives
\begin{equation}
3(f_4^2+f_5^2+f_6^2+f_7^2)=6f_2^2+f_3^2+2|\alpha|^2+2|\beta|^2+|\gamma|^2+|\delta|^2+|\zeta|^2\,,
\end{equation}
while the first $FF$-equation gives
\begin{equation}
5f_7^2+3f_5^2+3f_6^2+3|\zeta|^2-f_3^2+f_4^2+|\gamma|^2+|\delta|^2=2(-f_1^2+|\alpha|^2+|\beta|^2+f_2^2)
\end{equation}
giving
\begin{equation}
f_4^2=f_1^2+2f_2^2+f_7^2+|\gamma|^2+|\delta|^2+2|\zeta|^2
\end{equation}
and since $H\wedge F^{(4)}=0$ implies $f_1f_4=0$ we find $f_1=0$. We first analyze the case $f_3\neq0$. In this case we also get $\alpha=f_8\vartheta^1$ and $\beta=f_9\vartheta^1$. Without loss of generality we can take $\zeta=f_{10}\vartheta^{12}+f_{11}\vartheta^{23}$, $\gamma=\vartheta^1\wedge\gamma'+f_{12}\vartheta^{23}$ and $\delta=\vartheta^1\wedge\delta'+f_{13}\vartheta^{23}$ and the Einstein equation in the 1-direction gives
\begin{equation}
2f_8^2+2f_9^2+2f_2^2
-f_3^2
+|\gamma'|^2
+|\delta'|^2
+f_{10}^2
-(f_7^2+f_{11}^2+f_5^2+f_6^2+f_4^2+f_{12}^2+f_{13}^2)=0\,.
\end{equation}
Assume $f_8=f_9=0$. Then $H\wedge F^{(4)}=0$ implies $f_2f_4=0$ so that by the previous equation $f_2=0$ and $H$ vanishes. Using the above equations we find that all flux vanishes giving a contradiction. We may therefore assume that $f_8\neq0$ and therefore $f_7=0$. The condition $H\wedge F^{(2)}=0$ now gives 
\begin{equation}
f_6=-f_5f_9/f_8\,,\qquad
f_{11}=-f_2f_5/f_8\,,\qquad
f_2f_5f_9=0\,,
\end{equation}
while $H\wedge*F^{(4)}=0$ gives
\begin{equation}
\gamma'=-(f_9/f_8)\delta'\,.
\end{equation}
Taking $\delta'=f_{14}\vartheta^2+f_{15}\vartheta^3$ the remaining components of the Einstein equation in the flat directions give
\begin{equation}
f_{15}^2=f_8^2f_{10}^2/(f_8^2+f_9^2)+f_{14}^2\,,\quad
f_{14}f_{15}=f_{15}(f_8f_{13}-f_{12}f_9)=f_{10}f_{11}+f_{14}(f_{13}-f_{12}f_9/f_8)=0\,,
\end{equation}
which implies $f_{14}=0$ and $f_{10}(f_8f_{13}-f_{12}f_9)=f_{10}f_{11}=0$. The second $FF$-equation gives
\begin{equation}
f_5f_9f_{15}
=
-f_5f_{12}-f_6f_{13}+f_3f_4
=
-f_4f_6-f_{11}f_{12}+f_3f_{13}
=
-f_4f_5-f_{11}f_{13}+f_3f_{12}=0\,.
\end{equation}
Using the previous equations it is not hard to show that there is no solution unless $f_{10}=0$. Therefore $f_{10}=f_{15}=0$ so that $\gamma'=\delta'=0$ and our ansatz reduces to $H=(f_8\sigma_1+f_9\sigma_2+f_2\vartheta^{23})\wedge\vartheta^1$, $F^{(4)}=f_3\nu\wedge\vartheta^1+f_4\sigma_1\wedge\sigma_2+f_{12}\sigma_1\wedge\vartheta^{23}+f_{13}\sigma_2\wedge\vartheta^{23}$, $F^{(2)}=f_5\sigma_1+f_6\sigma_2+f_{11}\vartheta^{23}$. The remaining condition from $H\wedge F^{(4)}=0$ gives
\begin{equation}
f_2f_4+f_8f_{13}+f_9f_{12}=0
\end{equation}
and the earlier conditions become
\begin{align}
&f_2f_5f_9=0\,,\quad
f_6=-f_5f_9/f_8\,,\quad
f_{11}=-f_2f_5/f_8\,,\quad
f_3^2=f_5^2+f_6^2+f_{11}^2\,,
\nonumber\\
&f_4^2=2f_2^2+f_{12}^2+f_{13}^2+2f_{11}^2\,,\quad
f_8^2+f_9^2=f_5^2+f_6^2+2f_{11}^2+f_{12}^2+f_{13}^2\,.
\end{align}
The first four equations imply that $f_5\neq0$ since otherwise $f_3=0$ contrary to our assumption. With a bit of work one can show that there is no solution unless $f_2=0$ and one finds
\begin{align}
&f_2=f_{11}=0\,,\quad
f_5=\pm f_3f_8/\sqrt{f_3^2+f_4^2}\,,\quad
f_6=\mp f_3f_9/\sqrt{f_3^2+f_4^2}\,,\quad
f_{12}=\pm f_4f_8/\sqrt{f_3^2+f_4^2}\,,
\nonumber\\
&f_{13}=\mp f_4f_9/\sqrt{f_3^2+f_4^2}\,,\quad
f_8^2+f_9^2=f_3^2+f_4^2\,,
\end{align}
giving the solution (rescaling $f_8,f_9$ by $\sqrt{f_3^2+f_4^2}$)
\begin{align}
&\bm{AdS_3\times S^2\times S^2\times T^3}\,,\qquad
H=\sqrt{f_3^2+f_4^2}\Big(f_8\sigma_1+f_9\sigma_2\Big)\wedge\vartheta^1\,,
\label{eq:23a}
\\
&\qquad
F^{(4)}=f_3\nu\wedge\vartheta^1+f_4\sigma_1\wedge\sigma_2+f_4(f_8\sigma_1-f_9\sigma_2)\wedge\vartheta^{23}\,,\quad
F^{(2)}=f_3(f_8\sigma_1-f_9\sigma_2)\,,
\nonumber
\end{align}
with $f_8^2+f_9^2=1$. The curvatures are $-\frac12(f_3^2+f_4^2)$, $(f_3^2+f_4^2)f_8^2$ and $(f_3^2+f_4^2)f_9^2$ respectively.

It remains to analyze the case $f_3=0$. In this case the ansatz reduces to $H=\sigma_1\wedge\alpha+\sigma_2\wedge\beta+f_2\vartheta^{123}$, $F^{(4)}=f_4\sigma_1\wedge\sigma_2+\sigma_1\wedge\gamma+\sigma_2\wedge\delta$, $F^{(2)}=f_5\sigma_1+f_6\sigma_2+\zeta$ and $F^{(0)}=f_7$ with $\alpha,\beta\in\Omega^1(T^3)$ and $\gamma,\delta,\zeta\in\Omega^2(T^3)$. The trace of the Einstein equation in the flat directions and the first $FF$-equation in (\ref{eq:IIA}) give
\begin{equation}
2|\alpha|^2+2|\beta|^2=6f_7^2+3f_5^2+3f_6^2+5|\zeta|^2+2|\gamma|^2+2|\delta|^2\,,\qquad
f_4^2=2f_2^2+f_7^2+|\gamma|^2+|\delta|^2+2|\zeta|^2\,.
\end{equation}
It is easy to see that if $\alpha$ and $\beta$ both vanish there is no non-trivial solution. Therefore we may take $\alpha=f_8\vartheta^1$ with $f_8\neq0$ and therefore $f_7=0$. The condition $H\wedge F^{(2)}=0$ implies either\footnote{Note that in (i) the $\vartheta^3$-term can be assumed to vanish since this is forced if $f_9\neq0$ and can be accomplished by a rotation if $f_9=0$.}
\begin{align}
(i)&\qquad f_5=f_6=0\,,\qquad\zeta=f_9\vartheta^{12}\,,\qquad\beta=f_{10}\vartheta^1+f_{11}\vartheta^2\,,
\\
(ii)&\qquad\beta=-(f_6f_8/f_5)\vartheta^1\,,\qquad\zeta=f_9\vartheta^{12}+f_{10}\vartheta^{23}\,,\qquad f_2=-f_8f_{10}/f_5\,,\qquad f_6f_{10}=0\,.
\nonumber
\end{align}
However, in the second case there are no solutions as we will now show. If $f_6\neq0$ we get $f_{10}=f_2=0$ and the second $FF$-equation implies $f_4=0$, which in turn implies $\gamma=\delta=\zeta=0$, but that is inconsistent with the Einstein equation in the flat directions. We conclude that $f_6=0$ and from (ii) we get $\beta=0$ while the remaining $HF$ and $FF$-equations give $\gamma=0$ and $\delta=f_{11}\vartheta^{12}+f_{12}\vartheta^{13}+(f_{10}f_4/f_5)\vartheta^{23}$ with $(f_5^2+f_{10}^2)f_4+f_5f_9f_{11}=0$. The Einstein equation in the flat directions implies
\begin{equation}
f_{11}f_{12}=f_{10}f_{12}f_4=f_{10}(f_9+f_{11}f_4/f_5)=0\,,\quad
f_8^2+f_{12}^2-f_{10}^2(1+f_4^2/f_5^2)=0\,,\quad
f_{12}^2=f_9^2+f_{11}^2\,,
\end{equation}
giving $f_{11}=0$ which implies $f_4=0$ and therefore $f_{12}=f_{10}=f_9=0$ but then we find $f_8=0$ contradicting our assumptions.

It remains to analyze case (i) above. Writing $\gamma=f_{12}\vartheta^{12}+f_{13}\vartheta^{13}+f_{14}\vartheta^{23}$ and $\delta=f_{15}\vartheta^{12}+f_{16}\vartheta^{13}+f_{17}\vartheta^{23}$ the $HF$ and $FF$-equations give
\begin{align}
& f_9f_{15}=f_{11}f_{15}=0\,,\quad
f_{12}=-f_{10}f_{15}/f_8\,,\quad
f_{13}=(f_{10}f_{11}f_{14}+f_2f_4f_{11}-f_8f_{10}f_{16})/(f_8^2+f_{11}^2)\,,
\nonumber\\
&f_{17}=-(f_{10}f_{11}f_{16}+f_2f_4f_8+f_8f_{10}f_{14})/(f_8^2+f_{11}^2)\,.
\end{align}
The Einstein equation in the flat directions implies
\begin{align}
&f_{10}f_{11}+f_{13}f_{14}+f_{16}f_{17}=0\,,\quad
f_{12}f_{14}+f_{15}f_{17}=0\,,\quad
f_{12}f_{13}+f_{15}f_{16}=0\,,
\nonumber\\
&f_8^2+f_{10}^2-f_{11}^2+f_{13}^2-f_{14}^2+f_{16}^2-f_{17}^2=0\,,\quad
f_9^2+f_{11}^2+f_{12}^2-f_{13}^2+f_{15}^2-f_{16}^2=0\,.
\end{align}
It is easy to see that there is no solution with $f_{15}\neq0$ and therefore $f_{15}=f_{12}=f_9=0$. Noting that if $f_{11}\neq0$ the equations we want to solve may be written as
\begin{align}
&f_4^2=2f_2^2+f_8^2+f_{10}^2+f_{11}^2\,,\quad
f_{13}=f_{11}(f_2f_4+f_{10}(f_{14}-f_8f'_{16}))/(f_8^2+f_{11}^2)\,,\quad
\\
&f_{17}=-f_{10}f'_{16}-f_8(f_2f_4+f_{10}(f_{14}-f_8f'_{16}))/(f_8^2+f_{11}^2)\,,\quad
\\
&f_{10}(f_{16}'^2-1)
-(f_{14}-f_8f_{16}')(f_2f_4+f_{10}(f_{14}-f_8f'_{16}))/(f_8^2+f_{11}^2)
=0\,,
\\
&f_{16}'^2-1
+(f_2f_4+f_{10}(f_{14}-f_8f'_{16}))^2/(f_8^2+f_{11}^2)^2
=0\,,\quad
f_8^2+f_{10}^2=f_{14}^2+f_{17}^2\,,
\end{align}
we find three branches of solutions
\begin{align}
(i)&\qquad 
f_{11}=f_{13}=f_{16}=0\,,\quad
f_{14}=\Big(-f_2f_4f_{10}\pm f_8\sqrt{(f_8^2+f_{10}^2)^2-f_2^2f_4^2}\,\Big)/(f_8^2+f_{10}^2)\,,
\nonumber\\
&\qquad
f_{17}=-(f_2f_4+f_{10}f_{14})/f_8\,,\quad
f_4^2=2f_2^2+f_8^2+f_{10}^2\,,
%ok
\\
\nonumber\\
(ii)&\qquad
f_{13}=0\,,\quad
f_4^2=(2f_2^2+f_8^2+f_{11}^2)/(1-f_2^2/(2f_8)^2)\,,\quad
f_{10}=\mp f_2f_4/(2f_8)\,,
\nonumber\\
&\qquad f_{14}=\pm f_8\,,\quad
f_{16}=\mp f_{11}\,,\quad
f_{17}=\pm f_{10}\,,
%ok
\\
\nonumber\\
(iii)&\qquad
f_{11}\neq0\,,\quad
f_{16}=f_{11}f'_{16}\,,\quad
f_4^2=2f_2^2+f_8^2+f_{10}^2+f_{11}^2\,,\quad
f_{13}=f_2f_4f_{11}/(f_8^2+f_{11}^2+f_{10}^2)\,,
\nonumber\\
&\qquad
f_{14}=f_8f'_{16}-f_2f_4f_{10}/(f_8^2+f_{11}^2+f_{10}^2)\,,\quad
f_{17}=-f_{10}f'_{16}-f_2f_4f_8/(f_8^2+f_{11}^2+f_{10}^2)\,,
\nonumber\\
&\qquad
f_{16}'^2=1-f_2^2f_4^2/(f_8^2+f_{11}^2+f_{10}^2)^2\,.
%ok
\end{align}
Plugging this into our ansatz for the fluxes we get the solutions
\begin{align}
&\bm{AdS_3\times S^2\times S^2\times T^3}\,,\qquad
H=(f_8\sigma_1+f_{10}\sigma_2+f_2\vartheta^{23})\wedge\vartheta^1\,,
\nonumber\\
&\qquad
F^{(4)}=f_4\sigma_1\wedge\sigma_2+f_{14}\sigma_1\wedge\vartheta^{23}+f_{17}\sigma_2\wedge\vartheta^{23}\,,
\nonumber\\
&\qquad
f_4^2=2f_2^2+f_8^2+f_{10}^2\,,\quad
f_{14}=-\Big(f_2f_4f_{10}\pm f_8\sqrt{(f_8^2+f_{10}^2)^2-f_2^2f_4^2}\,\Big)/(f_8^2+f_{10}^2)\,,
\nonumber\\
&\qquad
f_{17}=-\Big(f_2f_4f_8\mp f_{10}\sqrt{(f_8^2+f_{10}^2)^2-f_2^2f_4^2}\,\Big)/(f_8^2+f_{10}^2)\,,
\label{eq:23b}
\end{align}
with curvatures $-\frac14(f_4^2+f_{14}^2+f_{17}^2)$, $\frac12(f_2^2+f_8^2+f_{14}^2)$ and $\frac12(f_2^2+f_{10}^2+f_{17}^2)$, this solution can be obtained by setting the curvature of one $S^2$ to zero in (\ref{eq:22}), and
\begin{align}
&\bm{AdS_3\times S^2\times S^2\times T^3}\,,\qquad
H=f_8\sigma_1\wedge\vartheta^1+\sigma_2\wedge(f_{10}\vartheta^1+f_{11}\vartheta^2)+f_2\vartheta^{123}\,,
\nonumber\\
&\qquad
F^{(4)}=f_4\sigma_1\wedge\sigma_2+f_8\sigma_1\wedge\vartheta^{23}-\sigma_2\wedge(f_{11}\vartheta^{13}-f_{10}\vartheta^{23})\,,\qquad
\nonumber\\
&\qquad
f_4^2=(2f_2^2+f_8^2+f_{11}^2)/(1-f_2^2/(2f_8)^2)\,,\quad
f_{10}=-f_2f_4/(2f_8)\,,
\label{eq:23c}
\end{align}
with curvatures $-\frac14(f_4^2+f_8^2+f_{10}^2+f_{11}^2)$, $\frac12(f_2^2+2f_8^2)$ and $\frac12(3f_2^4+8f_8^2f_{11}^2+6f_2^2f_8^2)/(4f_8^2-f_2^2)$ and
\begin{align}
&\bm{AdS_3\times S^2\times S^2\times T^3}\,,\qquad
H=f_8\sigma_1\wedge\vartheta^1+\sigma_2\wedge(f_{10}\vartheta^1+f_{11}\vartheta^2)+f_2\vartheta^{123}\,,
\nonumber\\
&\qquad
F^{(4)}=f_4\sigma_1\wedge\sigma_2+\sigma_1\wedge(f_{13}\vartheta^{13}+f_{14}\vartheta^{23})+\sigma_2\wedge(f_{11}f'_{16}\vartheta^{13}+f_{17}\vartheta^{23})\,,
\nonumber\\
&\qquad
f_{11}\neq0\,,\quad
f_4^2=2f_2^2+f_8^2+f_{10}^2+f_{11}^2\,,\quad
f_{13}=f_2f_4f_{11}/(f_8^2+f_{11}^2+f_{10}^2)\,,
\nonumber\\
&\qquad
f_{14}=f_8f'_{16}-f_2f_4f_{10}/(f_8^2+f_{11}^2+f_{10}^2)\,,\quad
f_{17}=-f_{10}f'_{16}-f_2f_4f_8/(f_8^2+f_{11}^2+f_{10}^2)\,,
\nonumber\\
&\qquad
f_{16}'^2=1-f_2^2f_4^2/(f_8^2+f_{11}^2+f_{10}^2)^2\,,
\label{eq:23d}
\end{align}
with curvatures $-\frac14(f_4^2+f_8^2+f_{10}^2+f_{11}^2)$, $\frac12(f_2^2+f_8^2+f_{13}^2+f_{14}^2)$ and $\frac12(f_2^2+f_{10}^2+f_{11}^2(1+f_{16}'^2)+f_{17}^2)$. All three branches lift to $D=11$ \cite{FigueroaO'Farrill:2011fj,Wulff:2016vqy}.

A special case of this analysis gives the solution
\begin{align}
&\bm{AdS_3\times\mathbbm{CP}^2\times T^3}\,,\qquad
H=f(\omega-\vartheta^{23})\wedge\vartheta^1\,,\quad
F^{(4)}=f\omega\wedge(\omega+\vartheta^{23})\,,
\end{align}
with curvatures $-\frac32f^2$ and $\frac32f^2$. This solution can be obtained by setting $f_2=f_3$ in (\ref{eq:16}).

For $AdS_3\times S^2\times T^5$ the flux takes the form $H=f_1\nu+\sigma\wedge\alpha+\beta$, $F^{(4)}=f_2\nu\wedge\vartheta^1+\sigma\wedge\gamma+\delta$, $F^{(2)}=f_3\sigma+\zeta$ and $F^{(0)}=f_4$ with $\alpha\in\Omega^1(T^5)$, $\beta\in\Omega^3(T^5)$, $\gamma,\zeta\in\Omega^2(T^3)$ and $\delta\in\Omega^4(T^5)$. Taking the trace of the Einstein equation in the flat directions gives
\begin{equation}
2|\alpha|^2
+6|\beta|^2
+3f_2^2
-|\gamma|^2
+3|\delta|^2
-|\zeta|^2
-5(f_3^2+f_4^2)
=0\,,
\end{equation}
while the first $FF$-equation gives
\begin{equation}
5f_4^2+3f_3^2+3|\zeta|^2-f_2^2+|\gamma|^2+|\delta|^2=2(-f_1^2+|\alpha|^2+|\beta|^2)\,,
\end{equation}
implying
\begin{equation}
f_3^2=f_1^2+f_2^2+2|\beta|^2+|\zeta|^2+2|\delta|^2\,.
\end{equation}
The condition $H\wedge F^{(2)}=0$ implies $f_1f_3=0$, so that $f_1=0$, as well as $f_3\beta=\alpha\wedge\zeta$ and $\beta\wedge\zeta=0$. The above equations give
\begin{equation}
2|\alpha|^2=5f_4^2+2f_2^2+4|\beta|^2+|\gamma|^2+7|\delta|^2+6|\zeta|^2\,,
\end{equation}
so clearly $\alpha$ cannot vanish. Therefore we must have $f_4=0$. If $f_2\neq0$ we must have $\alpha=f_5\vartheta^1$ and if $f_2=0$ we can also impose this since we are free to rotate the $T^5$ directions. The Einstein equation in the 1-direction reads
\begin{equation}
2f_5^2
+2|i_1\beta|^2
-f_2^2
+2|i_1\gamma|^2
+2|i_1\delta|^2
+2|i_1\zeta|^2
-(|\gamma|^2+|\delta|^2+f_3^2+|\zeta|^2)
=0\,.
\end{equation}
Using the above equations this forces $\beta=\delta=\zeta=0$ and, without loss of generality, $\gamma=f_6\vartheta^{23}+f_7\vartheta^{45}$. The remaining conditions from the $HF$ and $FF$-equations and flat Einstein equations give
\begin{equation}
f_3f_7+f_2f_6=0\,,\quad
f_3f_6+f_2f_7=0\,,\quad
f_3^2=f_2^2\,,\quad
f_5^2=f_2^2+f_6^2\,,\quad
f_7^2=f_6^2
\end{equation}
and we find the solution
\begin{equation}
\bm{AdS_3\times S^2\times T^5}\,,\qquad
H=f_5\sigma\wedge\vartheta^1\,,\quad
F^{(4)}=f_2\nu\wedge\vartheta^1+f_6\sigma\wedge(\vartheta^{23}\mp\vartheta^{45})\,,\quad
F^{(2)}=\pm f_2\sigma\,,
\label{eq:24}
\end{equation}
with $f_5^2=f_2^2+f_6^2$. The curvatures are $-\frac12f_5^2$ and $f_5^2$ respectively. If either $f_2$ or $f_6$ vanishes this solution lifts to $D=11$ \cite{FigueroaO'Farrill:2011fj,Wulff:2016vqy}.

\subsection{$AdS_3$ solutions: IIB}
For $AdS_3\times S^7$ the flux takes the form $H=f_1\nu$ and $F^{(3)}=f_2\nu$. The condition $H\wedge*F^{(3)}=0$ forces $f_1f_2=0$ but that is incompatible with the first $FF$-equation in (\ref{eq:IIB}). For $S^6\times S^1$ the flux is $H=f_1\nu$, $F^{(3)}=f_2\nu$ and $F^{(1)}=f_3\vartheta^1$. The Einstein equation in the flat direction gives $f_2^2+f_3^2=0$ and there is no solution. For $\mathbbm{CP}^3\times S^1$ or $G_{\mathbbm R}^+(2,5)\times S^1$ the flux takes the form $H=f_1\nu+f_2\omega\wedge\vartheta^1$, $F^{(5)}=f_3(1+*)\nu\wedge\omega$, $F^{(3)}=f_4\nu+f_5\omega\wedge\vartheta^1$ and $F^{(1)}=f_6\vartheta^1$. The Einstein equation in the flat direction gives $6f_2^2+3f_3^2+f_4^2+3f_5^2+f_6^2=0$ and again there is no solution.

For $AdS_3\times S^5\times S^2$ the flux takes the form $H=f_1\nu$, $F^{(5)}=f_2(1+*)\sigma_1$ and $F^{(3)}=f_3\nu$. The condition $H\wedge*F^{(3)}=0$ forces $f_1f_3=0$ but then the first $FF$-equation in (\ref{eq:IIB}) forces $f_1=f_3=0$ and we have the solutions
\begin{equation}
\bm{AdS_3\times S^5\times H^2}\,,\qquad\bm{AdS_3\times SLAG_3\times H^2}\,,\qquad
F^{(5)}=f_2(1+*)\sigma_1\,,
\label{eq:14-15}
\end{equation}
with curvatures $-\frac14f_2^2$, $\frac14f_2^2$ and $-\frac14f_2^2$ respectively. The former is T-dual to the special case of (\ref{eq:16}) with $f_2=0$. Replacing $S^2$ by $T^2$ the flux is the same but now with $F^{(1)}=f_4\vartheta^1$. However, the difference of the Einstein equation in the flat directions gives $f_4^2=0$ and we are back at the above ansatz.

The next possibility is $AdS_3\times S^4\times S^3$ with flux $H=f_1\nu+f_2\sigma_2$ and $F^{(3)}=f_3\nu+f_4\sigma_2$. The $HF$-equations give $f_1f_4-f_2f_3=0$ and $f_1f_3-f_2f_4=0$ while the first $FF$-equation gives $f_1^2+f_4^2=f_2^2+f_3^2$. The solutions are $f_1=\pm f_2$, $f_4=\pm f_3$ or $f_1=\mp f_2$, $f_3=-f_2$, $f_4=\pm f_2$ but the curvature of the $S^4$ is proportional to $f_2^2+f_4^2-f_1^2-f_3^2$ which vanishes and there is no solution.

For $AdS_3\times S^4\times S^2\times S^1$ the flux is $H=f_1\nu+f_2\sigma_2\wedge\vartheta^1$, $F^{(5)}=(1+*)f_3\sigma_1\wedge\vartheta^1$, $F^{(3)}=f_4\nu+f_5\sigma_2\wedge\vartheta^1$ and $F^{(1)}=f_6\vartheta^1$. The Einstein equation in the flat direction gives $2f_2^2+f_3^2+f_4^2+f_5^2+f_6^2=0$ and there is no solution.

For $AdS_3\times S^4\times T^3$ the flux is, without loss of generality, $H=f_1\nu+f_2\vartheta^{123}$, $F^{(4)}=f_3(1+*)\sigma\wedge\vartheta^1$, $F^{(3)}=f_4\nu+f_5\vartheta^{123}$ and $F^{(1)}=f_6\vartheta^1+f_7\vartheta^2$. The difference of the Einstein equation in the 2 and 3 direction forces $f_7=0$ while the difference in the 1 and 3 direction gives $f_3^2+2f_6^2=0$. The ansatz is then a special case of that analyzed above and there is again no solution.

For $AdS_3\times S^3\times S^3\times S^1$ the flux takes the form $H=f_1\nu+f_2\sigma_1+f_3\sigma_2$, $F^{(3)}=f_4\nu+f_5\sigma_1+f_6\sigma_2$ and $F^{(1)}=f_7\vartheta^1$. If $f_7\neq0$ we must have $H=0$ and the first $FF$-equation gives $f_4^2=2f_7^2+f_5^2+f_6^2$ while the Einstein equation in the flat direction gives $f_7^2+f_4^2=f_5^2+f_6^2$ which contradicts the assumption. Therefore $f_7=0$ and the first $FF$-equation gives $f_1^2+f_5^2+f_6^2=f_4^2+f_2^2+f_3^2$ while the Einstein equation in the flat direction gives $f_4^2=f_5^2+f_6^2$ so that $f_1^2=f_2^2+f_3^2$. The $HF$-equations now give $f_2f_6-f_3f_5=f_1f_5-f_2f_4=f_1f_6-f_3f_4=0$ and $-f_1f_4+f_2f_5+f_3f_6=0$ giving the solution (rescaling $f_{2,3}$ by $f_1$ and $f_{5,6}$ by $f_4$)
\begin{equation}
\bm{AdS_3\times S^3\times S^3\times S^1}\,,\qquad
H=f_1(\nu+f_2\sigma_1+f_3\sigma_2)\,,\quad
F^{(3)}=f_4(\nu+f_2\sigma_1+f_3\sigma_2)\,,\quad
f_2^2+f_3^2=1\,,
\end{equation}
with curvatures $-\frac12(f_1^2+f_4^2)$, $\frac12(f_1^2+f_4^2)f_2^2$ and $\frac12(f_1^2+f_4^2)f_3^2$ respectively. This well-known solution is T-dual to (\ref{eq:18}).

For $AdS_3\times S^3\times S^2\times S^2$ the flux takes the form $H=f_1\nu+f_2\sigma_1$, $F^{(5)}=f_3(1+*)\sigma_1\wedge\sigma_2+f_4(1+*)\sigma_1\wedge\sigma_3$ and $F^{(3)}=f_5\nu+f_6\sigma_1$. The $FF$ and $HF$-equations for $F^{(3)}$ give $f_1^2+f_6^2=f_5^2+f_2^2$, $f_1f_6-f_2f_5=0$ and $-f_1f_5+f_2f_6=0$ with solutions $f_1=\pm f_2$, $f_6=\pm f_5$ or $f_1=\mp f_2$, $f_6=\pm f_2$, $f_5=-f_2$. The remaining $HF$ and $FF$-equations give $f_1f_3-f_2f_4=f_1f_4-f_2f_3=0$ and $f_5f_3-f_6f_4=f_5f_4-f_6f_3=0$ giving the solutions
\begin{align}
(i)&\qquad
f_1=f_2=f_5=f_6=0\,,
\\
(ii)&\qquad
f_1=\pm f_2\,,\quad
f_6=\pm f_5\,,\quad
f_4=\pm f_3\,,
\\
(iii)&\qquad
f_1=\mp f_2\,,\quad
f_4=\mp f_3\,,\quad
f_5=-f_2\,,\quad
f_6=\pm f_2\,.
\end{align}
Only for the first one is it possible for both $S^2$ curvatures to be non-vanishing and we find the solution
\begin{equation}
\bm{AdS_3\times S^3\times S^2\times H^2}\,,\qquad
F^{(5)}=f_3(1+*)\sigma_1\wedge\sigma_2+f_4(1+*)\sigma_1\wedge\sigma_3\,,
\label{eq:19}
\end{equation}
with curvatures $-\frac14(f_3^2+f_4^2)$, $\frac14(f_3^2+f_4^2)$, $\frac14(f_3^2-f_4^2)$ and $\frac14(-f_3^2+f_4^2)$. This solution is T-dual to the special case of (\ref{eq:22}) with $f_3=f_4=f_8=0$.

For $AdS_3\times S^3\times S^2\times T^2$ the flux takes the form $H=f_1\nu+f_2\sigma_1+f_3\sigma_2\wedge\vartheta^1$, $F^{(5)}=f_4(1+*)\sigma_1\wedge\sigma_2+f_5(1+*)\sigma_1\wedge\vartheta^{12}$, $F^{(3)}=f_6\nu+f_7\sigma_1+\sigma_2\wedge(f_8\vartheta^1+f_9\vartheta^2)$ and $F^{(1)}=f_{10}\vartheta^1+f_{11}\vartheta^2$. The Einstein equation in the flat directions gives
\begin{equation}
f_3^2+f_8^2-f_9^2+f_{10}^2-f_{11}^2=0\,,\qquad
f_3^2-f_4^2+f_5^2-f_7^2+f_6^2=0\,,
\end{equation}
while the first $FF$-equation gives
\begin{equation}
2f_{10}^2+2f_{11}^2-f_6^2+f_7^2+f_8^2+f_9^2=-f_1^2+f_2^2+f_3^2\,.
\end{equation}
Assume $f_2\neq0$. Then we get from the $HF$-equations that $f_9=f_{10}=f_{11}=0$ and using the above equations also $f_3=f_8=0$ but then the curvature of the $S^2$ is proportional to $f_4^2-f_5^2-f_7^2+f_6^2$ which vanishes by the above equations.

It remains to analyze the case $f_2=0$. From the previous equations we have 
\begin{equation}
f_4^2=f_1^2+f_5^2+f_8^2+f_9^2+2f_{10}^2+2f_{11}^2
\end{equation}
and $f_4$ cannot vanish since then the curvature of the $S^2$ vanishes. The condition $H\wedge F^{(5)}=0$ implies $f_1f_4=f_1f_5=0$ so that $f_1=0$, while $H\wedge F^{(3)}=H\wedge *F^{(3)}=0$ implies $f_3f_6=f_3f_7=f_3f_8=0$. Assume $f_3=0$ so that $H=0$ and we may set also $f_9=0$ by a rotation. The remaining $FF$-equations gives $f_5f_7+f_4f_6=f_{10}f_8+f_4f_7+f_5f_6=0$ which together with the previous equations implies $f_8=f_{10}=f_{11}=0$ and vanishing $S^2$ curvature. Therefore we must have $f_3\neq0$ and $f_6=f_7=f_8=0$. The previous equations give $f_{10}=f_{11}=0$, $f_9^2=f_3^2$ and $f_4^2=f_3^2+f_5^2$ and the solution
\begin{align}
&\bm{AdS_3\times S^3\times S^2\times T^2}\,,\qquad
H=f_3\sigma_2\wedge\vartheta^1\,,\quad
F^{(5)}=f_4(1+*)\sigma_1\wedge\sigma_2+f_5(1+*)\sigma_1\wedge\vartheta^{12}\,,
\nonumber\\
&\qquad
F^{(3)}=f_3\sigma_2\wedge\vartheta^2\,,\qquad
f_4^2=f_3^2+f_5^2\,.
\end{align}
The curvatures are $-\frac12(f_3^2+f_5^2)$, $\frac12f_5^2$ and $f_3^2$ respectively. This solution is T-dual to the special case of (\ref{eq:18}) with $f_1=0$.

For $AdS_3\times S^3\times T^4$ the flux takes the form $H=f_1\nu+f_2\sigma+\alpha$, $F^{(5)}=(1+*)\sigma\wedge\beta$, $F^{(3)}=f_3\nu+f_4\sigma+\gamma$ and $F^{(1)}=f_5\vartheta^1$ where $\alpha,\gamma\in\Omega^3(T^4)$, $\beta\in\Omega^2(T^4)$. The trace of the Einstein equation in the flat directions gives
\begin{equation}
\tfrac32|\alpha|^2
+\tfrac12|\gamma|^2
-\tfrac12f_5^2
-f_4^2
+f_3^2
=0\,,
\end{equation}
while the first $FF$-equation gives
\begin{equation}
f_1^2+2f_5^2+f_4^2+|\gamma|^2=f_3^2+f_2^2+|\alpha|^2\,,
\end{equation}
implying $2f_2^2=2f_1^2+3f_5^2+|\alpha|^2+3|\gamma|^2$ and therefore $f_5=0$. If $f_2=0$ we get $f_1=\alpha=\gamma=0$ and $f_4^2=f_3^2$. While if $f_2\neq0$ we get from the $HF$-equations involving $F^{(3)}$  that $f_3=f_1f_4/f_2$, $\gamma=-(f_4/f_2)\alpha$ and $f_4(f_1^2-f_2^2+|\alpha|^2)=0$ implying $\alpha=\gamma=0$ and $f_2^2=f_1^2$. Writing $\beta=f_6\vartheta^{12}+f_7\vartheta^{34}$, without loss of generality, the remaining equations imply $f_7^2=f_6^2$ and $f_1(f_6\mp f_7)=0$ and we find the solution
\begin{equation}
\bm{AdS_3\times S^3\times T^4}\,,\qquad
H=f_1(\nu+\sigma)\,,\quad
F^{(5)}=f_6(1+*)\sigma\wedge(\vartheta^{12}+\vartheta^{34})\,,\quad
F^{(3)}=f_3(\nu+\sigma)\,,
\end{equation}
with curvatures $-\frac12(f_1^2+f_3^2+f_6^2)$ and $\frac12(f_1^2+f_3^2+f_6^2)$. This solution is T-dual to the one in (\ref{eq:21}).

For $AdS_3\times S^2\times S^2\times S^2\times S^1$ the flux takes the form $H=f_0\nu+f_i\sigma_i\wedge\vartheta^1$, $F^{(5)}=f_{ij}(1+*)\sigma_i\wedge\sigma_j\wedge\vartheta^1$, $F^{(3)}=g_0\nu+g_i\sigma_i\wedge\vartheta^1$ and $F^{(1)}=f_4\vartheta^1$. The Einstein equation in the flat direction gives $2|f_i|^2+|f_{ij}|^2+|g_i|^2+f_4^2+g_0^2=0$ and there is clearly no solution.

For $AdS_3\times S^2\times S^2\times T^3$ the flux takes the form $H=f_1\nu+\sigma_1\wedge\alpha+\sigma_2\wedge\beta+f_2\vartheta^{123}$, $F^{(5)}=(1+*)(f_3\sigma_1\wedge\sigma_2\wedge\vartheta^1+f_4\sigma_1\wedge\vartheta^{123}+f_5\sigma_2\wedge\vartheta^{123})$, $F^{(3)}=f_6\nu+\sigma_1\wedge\gamma+\sigma_2\wedge\delta+f_7\vartheta^{123}$ and $F^{(1)}=\zeta$ where $\alpha,\beta,\gamma,\delta,\zeta\in\Omega^1(T^3)$. The first $FF$-equation says that
\begin{equation}
f_1^2+2|\zeta|^2+|\gamma|^2+|\delta|^2+f_7^2=f_6^2+|\alpha|^2+|\beta|^2+f_2^2\,,
\end{equation}
while the trace of the Einstein equation in the flat directions gives
\begin{equation}
2|\alpha|^2+2|\beta|^2+6f_2^2+3f_4^2+3f_5^2+3f_6^2+3f_7^2=f_3^2+|\gamma|^2+|\delta|^2+|\zeta|^2\,,
\end{equation}
implying $f_3^2=2f_1^2+4f_2^2+3f_4^2+3f_5^2+f_6^2+5f_7^2+3|\zeta|^2+|\gamma|^2+|\delta|^2$ so that we must have $f_3\neq0$. The Einstein equation in the 1-direction reads
\begin{equation}
2f_2^2
+f_3^2+f_4^2+f_5^2
+f_6^2
+f_7^2
+2|i_1\alpha|^2
+2|i_1\beta|^2
+2|i_1\gamma|^2
+2|i_1\delta|^2
+2|i_1\zeta|^2
-|\gamma|^2
-|\delta|^2
-|\zeta|^2
=0
\end{equation}
and using the above equations we find $f_1=f_2=f_4=f_5=f_6=f_7=0$ and $\zeta=i_1\alpha=i_1\beta=i_1\gamma=i_1\delta=0$ and
\begin{equation}
f_3^2=|\alpha|^2+|\beta|^2=|\gamma|^2+|\delta|^2\,.
\end{equation}
Taking $\alpha=f_8\vartheta^2$, $\beta=f_9\vartheta^2+f_{10}\vartheta^3$, $\gamma=f_{11}\vartheta^2+f_{12}\vartheta^3$ and $\delta=f_{13}\vartheta^2+f_{14}\vartheta^3$ the remaining $HF$-equations and flat Einstein equations give
\begin{align}
&f_9f_{12}-f_{10}f_{11}+f_8f_{14}=0\,,\quad
f_9f_{13}+f_{10}f_{14}+f_8f_{11}=0\,,\quad
f_9f_{10}+f_{11}f_{12}+f_{13}f_{14}=0\,,
\nonumber\\
&f_8^2+f_9^2+f_{11}^2+f_{13}^2-f_{10}^2-f_{12}^2-f_{14}^2=0\,,
\end{align}
giving the two solutions
\begin{align}
(i)&\qquad
f_9=f_{14}=f_{11}=0\,,\quad
f_{12}=\pm f_8\,,\quad
f_{13}=\mp f_{10}\,,
\\
(ii)&\qquad
f_{11}=0\,,\quad
f_{12}=\pm f_8\,,\quad
f_{13}=\pm f_{10}\,,\quad
f_{14}=\mp f_9\,.
\end{align}
Plugging these into our ansatz for the fluxes we get the solutions
\begin{align}
&\bm{AdS_3\times S^2\times S^2\times T^3}\,,\qquad
H=f_8\sigma_1\wedge\vartheta^2+f_{10}\sigma_2\wedge\vartheta^3\,,\quad
F^{(5)}=f_3(1+*)\sigma_1\wedge\sigma_2\wedge\vartheta^1\,,
\nonumber\\
&\qquad
F^{(3)}=f_8\sigma_1\wedge\vartheta^3-f_{10}\sigma_2\wedge\vartheta^2\,,\qquad
f_3^2=f_8^2+f_{10}^2\,,
\label{eq:23e}
\end{align}
with curvatures $-\frac12f_3^2$, $f_8^2$ and $f_{10}^2$, and
\begin{align}
&\bm{AdS_3\times S^2\times S^2\times T^3}\,,\qquad
H=f_8\sigma_1\wedge\vartheta^2+\sigma_2\wedge(f_9\vartheta^2+f_{10}\vartheta^3)\,,\quad
F^{(5)}=f_3(1+*)\sigma_1\wedge\sigma_2\wedge\vartheta^1
\nonumber\\
&\qquad
F^{(3)}=f_8\sigma_1\wedge\vartheta^3+\sigma_2\wedge(f_{10}\vartheta^2-f_9\vartheta^3)\,,\qquad
f_3^2=f_8^2+f_9^2+f_{10}^2\,,
\label{eq:23f}
\end{align}
with curvatures $-\frac12f_3^2$, $f_8^2$ and $f_9^2+f_{10}^2$. These solutions are T-dual to the special case of (\ref{eq:23c}) and (\ref{eq:23d}) with $f_2=0$ respectively.

The analysis above also shows that there are no solutions with a $\mathbbm{CP}^2$ factor.

For $AdS_3\times S^2\times T^5$ the flux takes the form $H=f_1\nu+f_2\sigma\wedge\vartheta^1+\alpha$, $F^{(5)}=(1+*)\nu\wedge(\beta+f_3\sigma)$, $F^{(3)}=f_4\nu+\sigma\wedge\gamma+\delta$ and $F^{(1)}=\zeta$ where $\alpha,\delta\in\Omega^3(T^5)$, $\beta\in\Omega^2(T^5)$ and $\gamma,\zeta\in\Omega^1(T^5)$. The trace of the Einstein equation in the flat directions gives
\begin{equation}
2f_2^2
+6|\alpha|^2
+|\beta|^2
+5f_3^2
-3|\gamma|^2
+|\delta|^2
-3|\zeta|^2
+5f_4^2
=0\,,
\end{equation}
while the first $FF$-equation gives
\begin{equation}
f_1^2+2|\zeta|^2+|\gamma|^2+|\delta|^2=f_4^2+f_2^2+|\alpha|^2\,,
\end{equation}
leading to
\begin{align}
&f_2^2=3f_1^2+5f_3^2+2f_4^2+3|\alpha|^2+|\beta|^2+4|\delta|^2+3|\zeta|^2\,,
\nonumber\\
&|\gamma|^2=2f_1^2+5f_3^2+3f_4^2+4|\alpha|^2+|\beta|^2+3|\delta|^2+|\zeta|^2\,.
\end{align}
The Einstein equation in the 1-direction reads
\begin{equation}
2f_2^2
+2|i_1\alpha|^2
-2|i_1\beta|^2
+|\beta|^2
+f_3^2
+2|i_1\gamma|^2
+2|i_1\delta|^2
+2|i_1\zeta|^2
-|\gamma|^2
-|\delta|^2
-|\zeta|^2
+f_4^2
=0
\end{equation}
and using the previous equations this implies
\begin{equation}
2f_1^2
+3f_3^2
+f_4^2
+|\alpha|^2
+|\beta|^2
+2|\delta|^2
+2|\zeta|^2
+|i_1\alpha|^2
-|i_1\beta|^2
+|i_1\gamma|^2
+|i_1\delta|^2
+|i_1\zeta|^2
=0\,,
\end{equation}
so that $f_1=f_3=f_4=0$, $\alpha=\delta=\zeta=0$, $i_1\gamma=0$ and, without loss of generality, $\beta=f_5\vartheta^{12}$. The Einstein equation in the flat directions implies $\gamma=f_6\vartheta^2$ and we get the solution
\begin{equation}
\bm{AdS_3\times S^2\times T^5}\,,\qquad
H=f_2\sigma\wedge\vartheta^1\,,\quad
F^{(5)}=f_2(1+*)\nu\wedge\vartheta^{12}\,,\quad
F^{(3)}=f_2\sigma\wedge\vartheta^2\,.
\end{equation}
The curvatures are $-\frac12f_2^2$ and $f_2^2$. This solution is T-dual to that in (\ref{eq:24}).

\section{Supersymmetry analysis}\label{sec:susy}
We now wish to analyze which of these backgrounds preserve any supersymmetry. This analysis is simplified by observing that type IIA solutions with only $H$ and $F^{(4)}$ flux lift directly to {symmetric space solutions in eleven dimensions with an additional $S^1$-factor. It is not hard to see that furthermore the supersymmetry analysis is the same in eleven dimensions and therefore the supersymmetries of these backgrounds follows directly from the analysis performed in \cite{Wulff:2016vqy}.} The same applies to type IIB backgrounds T-dual to these IIA backgrounds on an $S^1$-factor.\footnote{It is not true for T-duality on for example the Hopf fiber of $S^3$ which can break some of the supersymmetries. {In general, for a T-duality on an $S^1$ (fibered or not), the Killing spinors that are preserved are the ones that do not depend on the $S^1$ coordinate. A fancier way to say this is that the preserved supersymmetries are determined by the so-called Kosmann derivative \cite{Kelekci:2014ima}.}} From that analysis it follows that the only supersymmetric ones are backgrounds 18 and 21 in table \ref{tab:AdS3} (some branches of solution 23 are also covered by that analysis but we will come back to this below).

It now remains only to analyze backgrounds 1, 2 and 4 in table \ref{tab:AdS5}, the backgrounds in table \ref{tab:AdS4} and backgrounds 14, 15, 19, 20, 23 and 24 in table \ref{tab:AdS3}. Solution 1 is $AdS_5\times S^5$ which is well known to be maximally supersymmetric. Solutions 20 and 24 are also well known supersymmetric near-horizon geometries \cite{Boonstra:1998yu,Wulff:2014kja}. We will analyze the remaining ones in turn below.

The conditions needed for the backgrounds to preserve some supersymmetries are the integrability of the Killing spinor equation and the dilatino equation. In the case of type IIA symmetric spaces they take the form (see \cite{Wulff:2013kga} whose conventions we follow)\footnote{Recall that all terms involving derivatives of background fields vanish in the case of symmetric spaces.}
\begin{align}
(R_{ab}{}^{cd}\Gamma_{cd}-\tfrac18G_{[a}G_{b]})\xi=&0\,,\qquad G_a=H_{abc}\Gamma^{bc}\Gamma_{11}+\mathcal S\Gamma_a\,,
\label{eq:KSP-int}
\\
(4\slashed H\Gamma_{11}+\Gamma^a\mathcal S\Gamma_a)\xi=&0\,.
\label{eq:dilatino}
\end{align}
In the type IIB case they take formally the same form provided one replaces the $32\times32$ gamma matrices $\Gamma_a$ by $16\times16$ ones $\gamma_a$ and $\Gamma_{11}$ by $\sigma^3$ acting on the two MW spinors of type IIB. In these equations the fluxes appear contracted with gamma matrices in the combinations $\slashed H=\frac16\Gamma^{abc}H_{abc}$ and
\begin{align}
\mathbf{IIA:}&\qquad
\mathcal S=F^{(0)}+\tfrac12F^{(2)}_{ab}\Gamma^{ab}\Gamma_{11}+\tfrac{1}{4!}F^{(4)}_{abcd}\Gamma^{abcd}\,,
\label{eq:SA}
\\
\mathbf{IIB:}&\qquad
\mathcal S=-(F^{(1)}_ai\sigma^2\gamma^a+\tfrac{1}{3!}F^{(3)}_{abc}\sigma^1\gamma^{abc}+\tfrac{1}{2\cdot5!}F^{(5)}_{abcde}i\sigma^2\gamma^{abcde})\,.
\label{eq:SB}
\end{align}
We will now analyze these equations for the backgrounds which remain after the observations above.

\subsubsection*{2. $AdS_5\times SLAG_3$}
Using the form of the fluxes in (\ref{eq:1-2}) in (\ref{eq:SB}) we find
\begin{equation}
\mathcal S=\tfrac12fi\sigma^2(\gamma^{01234}-\gamma^{56789})\,.
\end{equation}
Using this in the $(59)$-component of (\ref{eq:KSP-int}), together with the fact that $R_{59}{}^{ab}=0$ from (\ref{eq:SLAG3-R}), we get $f^2\xi=0$ and no non-trivial solution ruling out supersymmetry for this background.

\subsubsection*{4. $AdS_5\times S^3\times S^2$}
Using the form of the fluxes in (\ref{eq:4}) in (\ref{eq:SB}) we find
\begin{equation}
\mathcal S=\tfrac12f_2i\sigma^2(\gamma^{01234}-\gamma^{56789})\,.%=f_2\varepsilon\gamma^{01234}
\end{equation}
Using this in the $(59)$-component of (\ref{eq:KSP-int}) one finds again that there is no supersymmetric solution.

\subsubsection*{6. $AdS_4\times\mathbbm{CP}^3$}
Using the form of the fluxes in (\ref{eq:6-7}) in (\ref{eq:SA}) we find\footnote{{In the special case when the Romans mass vanishes the $AdS_4$ solutions with Riemannian factor being a K\"ahler space, i.e. $\mathbbm{CP}^3$, $\mathbbm{CP}^2\times S^2$ or $S^2\times S^2\times S^2$, lift to Sasaki-Einstein solutions in 11 dimensions. The Sasaki-Einstein space is a $U(1)$ fibration over the K\"ahler base and dimensional reduction on the $U(1)$ fiber typically breaks the supersymmetry completely \cite{Pope:1984jj,Duff:1986hr}. This is consistent with our findings here.}}
\begin{equation}
\mathcal S=f_4+f_3(\Gamma^{45}+\Gamma^{67}+\Gamma^{89})\Gamma_{11}-f_1\Gamma^{0123}+f_2(\Gamma^{4567}+\Gamma^{4589}+\Gamma^{6789})\,.
\label{eq:S6}
\end{equation}
The dilatino equation (\ref{eq:dilatino}) reads
\begin{equation}
\big[5f_4-3f_3(\Gamma^{45}+\Gamma^{67}+\Gamma^{89})\Gamma_{11}-f_1\Gamma^{0123}+f_2(\Gamma^{4567}+\Gamma^{4589}+\Gamma^{6789})\big]\xi=0\,.
\label{eq:6-dilatino}
\end{equation}
Using this the (04)-component of (\ref{eq:KSP-int}) gives
\begin{equation}
\big[3f_4-f_3(2\Gamma^{45}+\Gamma^{67}+\Gamma^{89})\Gamma_{11}+f_2\Gamma^{6789}\big]\big[f_4-f_3(\Gamma^{67}+\Gamma^{89})\Gamma_{11}+f_2\Gamma^{6789}\big]\xi=0\,.
\label{eq:6-04}
\end{equation}
Since the two matrix factors commute with each other we can analyze separately the case where the first(second) factor annihilates $\xi$. Assume that the first factor annihilates $\xi$. Multiplying with the same factor but with the sign of the $\Gamma^{45}$-term changed gives
\begin{equation}
\big[
f_2^2
+2f_3^2
+9f_4^2
+2(3f_2f_4+f_3^2)\Gamma^{6789}
+2(f_2-3f_4)f_3(\Gamma^{67}+\Gamma^{89})\Gamma_{11}
\big]\xi
=0\,.
\end{equation}
Again we multiply with the same factor but with the sign of the last term changed and we get
\begin{align}
\Big(
&(f_2^2+2f_3^2+9f_4^2)^2
+4(3f_2f_4+f_3^2)^2
+8(f_2-3f_4)^2f_3^2
\nonumber\\
&{}
+4\big[(f_2^2+2f_3^2+9f_4^2)(3f_2f_4+f_3^2)-2(f_2-3f_4)^2f_3^2\big]\Gamma^{6789}
\Big)\xi
=0\,.
\end{align}
Since the eigenvalues of $\Gamma^{6789}$ are $\pm1$ we find that for any supersymmetry to be preserved we need
\begin{equation}
[f_2^2+2f_3^2+9f_4^2\mp2(3f_2f_4+f_3^2)]^2+8(1\pm1)(f_2-3f_4)^2f_3^2=0\,,
\end{equation}
and using the relations between the parameter in (\ref{eq:6-7}) it is easy to see that we must have (note that $f_1^2\geq 9f_3^2$)
\begin{equation}
f_2=f_4=0\,,\qquad
f_1^2=9f_3^2\,,
%(3f_3\Gamma^{89}\Gamma_{11}+f_1\Gamma^{0123})\xi=0
\end{equation}
with $(1+\Gamma^{6789})\xi=(1-\Gamma^{4567})\xi=0$. In the other case, when the second factor in (\ref{eq:6-04}) annihilates $\xi$, one finds again the same conditions on the parameters but now with $(1-\Gamma^{6789})\xi=0$.
%%%%
\iffalse
Assume $(f_4-f_3(\Gamma^{67}+\Gamma^{89})\Gamma_{11}+f_2\Gamma^{6789})\xi=0$ mul with same with sign of second term changed gives
$
(
f_4^2+f_2^2+2f_3^2
+2(f_4f_2-f_3^2)\Gamma^{6789}
)\xi=0
$
mul with same with sign of gamma-term changed implies, since $f_3$ cannot vanish due to the constr, that $f_4=-f_2$ and then the constraints imply $f_2=f_4=0$ and $f_1=\pm 3f_3$. The dilatino eq then gives
$
(1\mp\Gamma^{012345}\Gamma_{11})\xi=0
$
while
06,08 satisfied!
\fi
%%%%
We therefore conclude that for supersymmetry we need\footnote{The dilatino equation (\ref{eq:6-dilatino}) implies that the solution with $f_1=+3f_3$ is non-supersymmetric.}
\begin{equation}
f_2=f_4=0\,,\qquad f_1=-3f_3\,,\qquad
(1-\Gamma^{4567}-\Gamma^{4589}-\Gamma^{6789})\xi=0\,.
\label{eq:cond6}
\end{equation}
This is compatible with the components of (\ref{eq:KSP-int}) with one index from $AdS_4$ and one from $\mathbbm{CP}^3$. Using the form of the $AdS$ curvature in (\ref{eq:AdS-S-R}) with $1/R^2=f_3^2$ one sees that the components of (\ref{eq:KSP-int}) with two $AdS$ indices are satisfied.

Finally, using the form of the $\mathbbm{CP}^3$ curvature in (\ref{eq:CP3-R}) with $1/R^2=f_3^2/4$, it is not hard to show that the remaining components of (\ref{eq:KSP-int}) are satisfied. The projection on $\xi$ removes eight components and this solution therefore preserves 24 supersymmetries. This solution is well known and arises by dimensional reduction on the Hopf fiber from the maximally supersymmetric $AdS_4\times S^7$ solution in $D=11$ \cite{Nilsson:1984bj}.

\subsubsection*{7. $AdS_4\times G_{\mathbbm R}^+(2,5)$}
The analysis is the same as for the previous background except for the last part where the curvature of $G_{\mathbbm R}^+(2,5)$ must be used in place of that of $\mathbbm{CP}^3$. Using the curvature in (\ref{eq:GR25-R}) with $1/R^2=f_3^2/4$ together with (\ref{eq:S6}) and (\ref{eq:cond6}) in the (45)-component of (\ref{eq:KSP-int}) one finds that there is no supersymmetric solution in this case.

\subsubsection*{8. $AdS_4\times S^4\times S^2$}
Using the form of the fluxes in (\ref{eq:8}) in (\ref{eq:SA}) we find
\begin{equation}
\mathcal S=f_4+f_3\Gamma^{89}\Gamma_{11}-f_1\Gamma^{0123}+f_2\Gamma^{4567}\,.
\end{equation}
The dilatino equation (\ref{eq:dilatino}) reads
\begin{equation}
(5f_4-3f_3\Gamma^{89}\Gamma_{11}-f_1\Gamma^{0123}+f_2\Gamma^{4567})\xi=0\,.
\end{equation}
Using this in the $(04)$-component of (\ref{eq:KSP-int}) gives
\begin{equation}
(3f_4^2-f_3^2-4f_3f_4\Gamma^{89}\Gamma_{11})\xi=0
\end{equation}
and multiplying with the same factor but with the sign of the last term changed gives $[(3f_4^2-f_3^2)^2+16f_3^2f_4^2]\xi=0$ implying $f_3=f_4=0$ but then all flux would vanish which rules out supersymmetry for this background.

\subsubsection*{9. $AdS_4\times\mathbbm{CP}^2\times S^2$}
Using the form of the fluxes in (\ref{eq:9}) in (\ref{eq:SA}) we find
\begin{equation}
\mathcal S=f_4'+f_0(f_1\Gamma^{45}+f_1\Gamma^{67}+f_3\Gamma^{89})\Gamma_{11}-f_0\Gamma^{0123}+f_{12}\Gamma^{4567}+f_{13}\Gamma^{4589}+f_{13}\Gamma^{6789}\,,
\end{equation}
where $f_4'=f_4(1-f_3-2f_1^2)$. The dilatino equation (\ref{eq:dilatino}) reads
\begin{equation}
\big(5f_4'-3f_0(f_1\Gamma^{45}+f_1\Gamma^{67}+f_3\Gamma^{89})\Gamma_{11}-f_0\Gamma^{0123}+f_{12}\Gamma^{4567}+f_{13}\Gamma^{4589}+f_{13}\Gamma^{6789}\big)\xi=0\,.
\label{eq:9-dilatino}
\end{equation}
Using this in the $(04)$-component of (\ref{eq:KSP-int}) we find
\begin{equation}
\big[3f_4'-f_0(2f_1\Gamma^{45}+f_1\Gamma^{67}+f_3\Gamma^{89})\Gamma_{11}+f_{13}\Gamma^{6789}\big]\big[f_4'-f_0(f_1\Gamma^{67}+f_3\Gamma^{89})\Gamma_{11}+f_{13}\Gamma^{6789}\big]\xi=0\,.
\label{eq:9-04}
\end{equation}
Since the two matrix factors commute we can analyze separately the case when the first(second) factor annihilates $\xi$. Assume that the second factor annihilates $\xi$. Multiplying with the same factor
 but with the sign of the second term changed gives, using $f_{13}=f_4f_1(1+f_3)$, the condition
\begin{equation}
\Big(
f_4^2(1-f_3-2f_1^2)^2
+f_4^2f_1^2(1+f_3)^2
+f_0^2(f_1^2+f_3^2)
+2f_1[f_4^2(1-f_3-2f_1^2)(1+f_3)-f_0^2f_3]\Gamma^{6789}
\Big)\xi
=0\,.
\end{equation}
Since the eigenvalues of $\Gamma^{6789}$ are $\pm1$ we find
\begin{equation}
f_4^2[(1-f_3-2f_1^2)\pm f_1(1+f_3)]^2+f_0^2(f_1\mp f_3)^2=0
\end{equation}
and since $f_0$ cannot vanish due to the relations between the parameters in (\ref{eq:9}) we get $f_3=\pm f_1$ and $f_4=0$ ($f_1^2=1$ is not compatible with (\ref{eq:9})). Using the relations in (\ref{eq:9}) this further implies the $f_{12}=f_{13}=0$. But then the $(48)$-component of (\ref{eq:KSP-int}) implies
\begin{equation}
(1+f_1\Gamma^{6789})\xi=0\,,
\end{equation}
but since $f_1^2=1$ is not allowed there is no non-trivial solution. In a similar way one rules out the second factor in the $(08)$-component of (\ref{eq:KSP-int}), written in a similar form to (\ref{eq:9-04}), annihilating $\xi$.
%%%%%%%
\iffalse
08:
$
(3f_4'-f_0(f_1\Gamma^{45}+f_1\Gamma^{67}+2f_3\Gamma^{89})\Gamma_{11}+f_{12}\Gamma^{4567})(f_4'-f_0(f_1\Gamma^{45}+f_1\Gamma^{67})\Gamma_{11}+f_{12}\Gamma^{4567})\xi=0
$
assume 2nd factor gives zero. mul with same with last term changed
$
f_4'^2
-f_{12}^2
-2f_0^2f_1^2
-2f_4'f_0(f_1\Gamma^{45}+f_1\Gamma^{67})\Gamma_{11}
+2f_0^2f_1^2\Gamma^{4567}
=0
$
again
$
(f_4'^2
-f_{12}^2
-2f_0^2f_1^2
+2f_0^2f_1^2\Gamma^{4567})^2
+4f_4'^2f_0^2f_1^2(1-\Gamma^{4567})^2
$
since the eigenvalues are $\pm1$ we find, using the constraints, that $f_4=0=f_{12}=f_{13}$ and $1-3[2f_1^2+f_3^2]=0$

48: gives $f_3^1=1$ and no solution
\fi
%%%%%%%

Therefore we must have first factor giving zero in both cases, i.e.
\begin{align}
\big[3f_4'-f_0(2f_1\Gamma^{45}+f_1\Gamma^{67}+f_3\Gamma^{89})\Gamma_{11}+f_{13}\Gamma^{6789}\big]\xi=&0\,,\\
\big[3f_4'-f_0(f_1\Gamma^{45}+f_1\Gamma^{67}+2f_3\Gamma^{89})\Gamma_{11}+f_{12}\Gamma^{4567}\big]\xi=&0\,.
\end{align}
Taking the difference of these two equations we get
\begin{equation}
\big[-f_0(f_1\Gamma^{45}-f_3\Gamma^{89})\Gamma_{11}+(f_{13}\Gamma^{89}-f_{12}\Gamma^{45})\Gamma^{67}]\xi=0\,.
\end{equation}
Multiplying with same expression but with the sign of the last term changed gives
\begin{equation}
\big[f_0^2(f_1+f_3\Gamma^{4589})^2+(f_{12}+f_{13}\Gamma^{4589})^2\big]\xi=0
\end{equation}
and since the eigenvalues of $\Gamma^{4589}$ are $\pm1$ this gives $f_3=\pm f_1$ and $f_{13}=\pm f_{12}$. Using this in the previous equations gives
\begin{equation}
\big[3f_4'-f_0(3f_1\Gamma^{45}+f_1\Gamma^{67})\Gamma_{11}+f_{12}\Gamma^{4567}]\xi=0\,.
\end{equation}
Multiplying with same expression but with the sign of the second term changed gives
\begin{equation}
\big[(3f_4'+f_{12}\Gamma^{4567})^2+f_0^2f_1^2(3-\Gamma^{4567})^2\big]\xi=0
\end{equation}
and since the eigenvalues are $\pm1$ this gives $f_1=0$ and $f_{12}=\pm3f_4'$ but since $f_3=\pm f_1=0$ one finds from the relations between the parameters in (\ref{eq:9}) that there is no non-trivial solution and supersymmetry is ruled out for this background.

\subsubsection*{10. $AdS_4\times S^3\times S^3$}
Using the form of the fluxes in (\ref{eq:10}) in (\ref{eq:SA}) we find $\mathcal S=f(1-\sqrt5\Gamma^{0123})$ and the dilatino equation (\ref{eq:dilatino}) gives $f(5-\sqrt5\Gamma^{0123})\xi=0$ which rules out supersymmetry.

\subsubsection*{11. $AdS_4\times S^3\times S^2\times S^1$}
Using the form of the fluxes in (\ref{eq:11}) in (\ref{eq:SA}) we find
\begin{equation}
\mathcal S=-f\varepsilon\gamma^9(\pm1+\frac{\sqrt5}{2}(\gamma^{0123}-\gamma^{456789}))\,,\qquad\slashed H=\sqrt2f\gamma^{789}\,.
\end{equation}
The dilatino equation (\ref{eq:dilatino}) reads $f(2\pm\sqrt2\sigma^1\gamma^{789})\xi=0$ which clearly has no non-trivial solution.

\subsubsection*{12. $AdS_4\times S^2\times S^2\times S^2$}
Using the form of the fluxes in (\ref{eq:12}) in (\ref{eq:SA}) we find
\begin{equation}
\mathcal S=f_4+f_0(f_1\Gamma^{45}+f_2\Gamma^{67}+f_3\Gamma^{89})\Gamma_{11}-f_0\Gamma^{0123}+f_{12}\Gamma^{4567}+f_{13}\Gamma^{4589}+f_{23}\Gamma^{6789}\,.
\end{equation}
The dilatino equation (\ref{eq:dilatino}) reads
\begin{equation}
\big(5f_4-3f_0(f_1\Gamma^{45}+f_2\Gamma^{67}+f_3\Gamma^{89})\Gamma_{11}-f_0\Gamma^{0123}+f_{12}\Gamma^{4567}+f_{13}\Gamma^{4589}+f_{23}\Gamma^{6789}\big)\xi=0\,.
\end{equation}
Using this in the $(04)$-component of (\ref{eq:KSP-int}) we get
\begin{equation}
\big(3f_4-f_0(2f_1\Gamma^{45}+f_2\Gamma^{67}+f_3\Gamma^{89})\Gamma_{11}+f_{23}\Gamma^{6789}\big)\big(f_4-f_0(f_2\Gamma^{67}+f_3\Gamma^{89})\Gamma_{11}+f_{23}\Gamma^{6789}\big)\xi=0\,.
\label{eq:04-12}
\end{equation}
The two matrix factors commute and we can analyze separately the case where the first(second) factor annihilates $\xi$. Assume the second factor annihilates $\xi$. Multiplying with the same factor but with the sign of the last term changed gives
\begin{equation}
\Big(
f_4^2
-f_0^2f_2^2
-f_0^2f_3^2
-f_{23}^2
+2f_0^2f_2f_3\Gamma^{6789}
-2f_0f_4(f_2\Gamma^{67}+f_3\Gamma^{89})\Gamma_{11}
\Big)\xi
=0\,.
\end{equation}
Doing the same again gives
\begin{equation}
\Big(
\big[f_4^2
-f_0^2f_2^2
-f_0^2f_3^2
-f_{23}^2
+2f_0^2f_2f_3\Gamma^{6789}
\big]^2
+4f_0^2f_4^2(f_2-f_3\Gamma^{6789})^2
\Big)\xi
=0\,.
\end{equation}
Since the eigenvalues of $\Gamma^{6789}$ are $\pm1$ this implies, noting that $f_{23}$ is proportional to $f_4$, that $f_3=\pm f_2$, $f_2(1\mp\Gamma^{6789})\xi=0$ and $f_{23}^2=f_4^2$. Using further the relations between the parameters in (\ref{eq:12}), in particular the fact that $f_1^2+f_2^2+f_3^2\leq\frac13$, the only solution is $f_4=f_{ij}=0$. Using this in the $(06)$-component of (\ref{eq:KSP-int}) gives $f_1^2=f_2^2=f_3^2=\frac19$ but then it is not hard to show that the $(68)$-component of the same equation has no solution.

Due to the symmetry in the $S^2$ factors we must then have that the first factor annihilates $\xi$ in the $(04)$, $(06)$ and $(08)$-component of (\ref{eq:KSP-int}), i.e., looking at (\ref{eq:04-12}),
\begin{align}
\big(3f_4-f_0(2f_1\Gamma^{45}+f_2\Gamma^{67}+f_3\Gamma^{89})\Gamma_{11}+f_{23}\Gamma^{6789}\big)\xi=&\,0\,,\\
\big(3f_4-f_0(f_1\Gamma^{45}+2f_2\Gamma^{67}+f_3\Gamma^{89})\Gamma_{11}+f_{13}\Gamma^{4589}\big)\xi=&\,0\,,\\
\big(3f_4-f_0(f_1\Gamma^{45}+f_2\Gamma^{67}+2f_3\Gamma^{89})\Gamma_{11}+f_{12}\Gamma^{4567}\big)\xi=&\,0\,.
\end{align}
Taking the difference of the first two equations we get
\begin{equation}
\big(f_0(f_1+f_2\Gamma^{4567})+(f_{13}+f_{23}\Gamma^{4567})\Gamma^{89}\Gamma_{11}\big)\xi=0\,.
\end{equation}
Multiplying with same factor but with the sign of the last term changed gives
\begin{equation}
\big(f_0^2(f_1+f_2\Gamma^{4567})^2+(f_{13}+f_{23}\Gamma^{4567})^2\big)\xi=0\,,
\end{equation}
implying $f_2^2=f_1^2$ and $f_{23}^2=f_{13}^2$ and $(f_1+f_2\Gamma^{4567})\xi=0$. In a similar way we find, by taking the difference of two equations, that $f_3^2=f_1^2$ and $f_{23}^2=f_{12}^2$ and $(f_1+f_3\Gamma^{4589})\xi=(f_2+f_3\Gamma^{6789})\xi=0$ and we are left with the single equation
\begin{equation}
\big(3f_4-4f_0f_1\Gamma^{45}\Gamma_{11}+f_{12}\Gamma^{4567}\big)\xi=0\,.
\end{equation}
Multiplying with the same expression but with the sign of the last term changed gives
\begin{equation}
\big(
9f_4^2
-24f_0f_1f_4\Gamma^{45}\Gamma_{11}
-16f_0^2f_1^2
-f_{12}^2
\big)\xi=0
\end{equation}
and it is easy to see, using the relations between parameters in (\ref{eq:12}), that there is no non-trivial solution. This rules out supersymmetry for this background.

\subsubsection*{13. $AdS_4\times S^2\times S^2\times T^2$}
Using the form of the fluxes in (\ref{eq:13}) in (\ref{eq:SA}) we find
\begin{equation}
\mathcal S=f_3(-\Gamma^{89}\Gamma_{11}+\sqrt5\Gamma^{4567})\,,\qquad\slashed H=\sqrt2f_3(\Gamma^{458}\pm\Gamma^{679})\,.
\end{equation}
The dilatino equation (\ref{eq:dilatino}) reads
\begin{equation}
f_3\Big(2\sqrt2(\Gamma^{458}\pm\Gamma^{679})\Gamma_{11}+3\Gamma^{89}\Gamma_{11}+\sqrt5\Gamma^{4567}\Big)\xi=0\,.
\end{equation}
Multiplying with the same expression but with the sign of last term changed gives $f_3\big(1+6\sqrt2(\Gamma^{459}\mp\Gamma^{678})\big)\xi=0$ which obviously has no non-trivial solution.

\subsubsection*{14 \& 15. $AdS_3\times S^5\times H^2$ and $AdS_3\times SLAG_3\times H^2$}
Using the form of the fluxes in (\ref{eq:14-15}) in (\ref{eq:SB}) we find
\begin{equation}
\mathcal S=\tfrac12f_2i\sigma^2(\gamma^{01289}-\gamma^{34567})\,.%=f_2\varepsilon\gamma^{01234}
\end{equation}
Using this in the $(29)$-component of (\ref{eq:KSP-int}) one finds again that there is no supersymmetric solution.

\subsubsection*{19. $AdS_3\times S^3\times S^2\times H^2$}
Using the form of the fluxes in (\ref{eq:19}) in (\ref{eq:SB}) we find $\mathcal S=-\varepsilon\gamma^{34567}(f_3-f_4\gamma^{6789})$. The $(68)$-component of (\ref{eq:KSP-int}) implies $(f_3^2-f_4^2)\xi=0$ but $f_3^2=f_4^2$ reduces the solution to $AdS_3\times S^3\times T^4$.

\subsubsection*{23. $AdS_3\times S^2\times S^2\times T^3$}
The last three branches of this solution, (\ref{eq:23b})--(\ref{eq:23d}), lift to $D=11$ and are therefore covered by the analysis in \cite{Wulff:2016vqy}. That analysis implies that supersymmetry requires $f_2=0$ and one further finds that the first two of these branches are non-supersymmetric while the third one preserves 8 supersymmetries provided that $f_2=f_{10}=0$ and $f_{14}=-f_8$ in (\ref{eq:23d}) leaving a one-parameter family of solutions which is also a well known near-horizon geometry \cite{Boonstra:1998yu,Wulff:2014kja}.

It remains to analyze the first branch (\ref{eq:23a}), which does not lift to $D=11$. Using the form of the fluxes in (\ref{eq:23a}) in (\ref{eq:SA}) we find
\begin{equation}
\mathcal S=f_3(f_8\Gamma^{34}-f_9\Gamma^{56})\Gamma_{11}+f_3\Gamma^{0127}+f_4\Gamma^{3456}+f_4f_8\Gamma^{3489}-f_4f_9\Gamma^{5689}\,,
\end{equation}
with $f_8^2+f_9^2=1$ and non-zero $\slashed H$ whose form we will not need here. The $(08)$-component of (\ref{eq:KSP-int}) reads
\begin{equation}
\big[f_3(f_8\Gamma^{34}-f_9\Gamma^{56})\Gamma_{11}-f_3\Gamma^{0127}+f_4\Gamma^{3456}-f_4f_8\Gamma^{3489}+f_4f_9\Gamma^{5689}\big]\big[f_4+f_3(f_8\Gamma^{56}-f_9\Gamma^{34})\Gamma_{11}\big]\xi=0\,.
\label{eq:08-23}
\end{equation}
The two matrix factors commute and we can analyze separately the case where the first(second) factor annihilates $\xi$. Assume the second factor annihilates $\xi$. Multiplying with the same factor but with the sign of the second term changed gives
$
\big(f_3^2+f_4^2+2f_3^2f_8f_9\Gamma^{3456}\big)\xi=0
$
and since the eigenvalues of $\Gamma^{3456}$ are $\pm1$ we find $f_3^2(f_8\pm f_9)^2+f_4^2=0$ and since $f_3$ and $f_4$ cannot both vanish we get $f_4=0$ and $f_9=\mp f_8$. But then the $(89)$-component of (\ref{eq:KSP-int}) says that $f_8f_9=0$ and there is no non-trivial solution.

Therefore the first factor in (\ref{eq:08-23}) must annihilate $\xi$, i.e.
\begin{equation}
\big[f_3(f_8\Gamma^{34}-f_9\Gamma^{56})\Gamma_{11}-f_3\Gamma^{0127}+f_4\Gamma^{3456}-f_4f_8\Gamma^{3489}+f_4f_9\Gamma^{5689}\big]\xi=0\,.
\end{equation}
The $(89)$-component of (\ref{eq:KSP-int}) now gives
\begin{equation}
f_4\big[f_3\Gamma^{0127}+f_4f_8\Gamma^{3489}-f_4f_9\Gamma^{5689}\big]\big[1+f_8\Gamma^{5689}-f_9\Gamma^{3489}\big]\xi=0\,.
\end{equation}
If $f_4=0$ we find from the previous equation that $f_8f_9=0$ but then the curvature of one of the $S^2$'s vanishes. Therefore $f_4\neq0$ and one of the other two factors gives zero. If the first of them does the previous equation reduces to the case analyzed before. We conclude that $(1+f_8\Gamma^{5689}-f_9\Gamma^{3489})\xi=0$ but this again implies $f_8f_9=0$ and vanishing curvature of one $S^2$. This rules out supersymmetry for this branch of the solution.

\section{Conclusions}
We have found all symmetric $AdS_n$ solutions of type II supergravity for $n>2$ and analyzed their supersymmetry. All supersymmetric solutions are well known already and come from near-horizon geometries of branes. An obvious and important extension of our results would be to find also all $AdS_2$ solutions. The analysis of these is quite involved due to the many invariant forms that many of these geometries admit, for some partial results in the type IIB case see \cite{FigueroaO'Farrill:2012rf}. {One can also obtain several $AdS_2$ solutions by double analytic continuation of the solutions with $S^2$ factors found in this paper. The only cases where this does not lead to real solutions are backgrounds 11 and 13 in table \ref{tab:AdS4}. We hope to address the full classification in the future.}

An interesting application of this classification of solutions is to the classification of integrable strings. The supersymmetric solutions we have found are known to lead to integrable superstrings \cite{Wulff:2014kja,Wulff:2015mwa} and an interesting question is whether any of the non-supersymmetric solutions can lead to integrable superstrings.\footnote{For solutions without NSNS flux the bosonic string is a symmetric space sigma model which is a standard example of an integrable system. In this case the inclusion of the fermions may or may not spoil the integrability.} This question was analyzed recently for the richest solution (in terms of free parameters) found here, $AdS_3\times S^2\times S^2\times T^3$ in table \ref{tab:AdS3}, in \cite{Wulff:2017hzy} where it was found that the superstring is integrable precisely for choices of the flux parameters such that the model is T-dual {(e.g. \cite{Lozano:2015bra})} to the supersymmetric {$AdS_3\times S^3\times S^3\times S^1$} solution. This analysis and other arguments suggest that generically one should only expect integrability in the supersymmetric case (or cases related by T-duality) but a proof is still lacking. Although this question is different from that of stability of the background analyzed in \cite{Ooguri:2016pdq,Freivogel:2016qwc} they both point towards supersymmetry playing a very important role in AdS/CFT. I plan to address the question of integrability for the remaining backgrounds found here in the near future.
	
\vspace{1cm}

\section*{{Acknowledgments}}
I thank the Galileo Galilei Institute for Theoretical Physics (GGI) for the hospitality and INFN for partial support during the completion of this work, within the program “New Developments in AdS3/CFT2 Holography”.

\vspace{3cm}

\appendix

\section{Riemann tensor for some symmetric spaces}
For the supersymmetry analysis we need the form of the Riemann tensor for some of the symmetric spaces appearing. Besides $AdS$ and spheres for which the Riemann tensor takes the well known form (in our conventions)
\begin{equation}
AdS_n:\quad R_{ab}{}^{cd}=\frac{2}{R^2}\delta_{[a}^c\delta_{b]}^d\,,\qquad S^n:\quad R_{ab}{}^{cd}=-\frac{2}{R^2}\delta_{[a}^c\delta_{b]}^d\,,
\label{eq:AdS-S-R}
\end{equation}
with $R$ the radius of curvature of the respective space, we need also the Riemann tensor for $\mathbbm{CP}^3$, $G_{\mathbbm R}^+(2,5)$ and $SLAG_3$. Since the expressions for these are less standard we will derive them here directly from the form of the isometry algebra.

The isometry algebra of an $n$-dimensional Riemannian symmetric space $G/H$ can be written in terms of generators $M_{ij}$ and $P_i$ ($i,j=1,\ldots n$) satisfying (see for example \cite{Wulff:2015mwa})
\begin{equation}
[M_{ij},M_{kl}]=\delta_{ik}M_{jl}-(i\leftrightarrow j)-(k\leftrightarrow l)\,,\qquad
[M_{ij},P_k]=2\delta_{k[i}P_{j]}\,,\qquad
[P_i,P_j]=-\tfrac12R_{ij}{}^{kl}M_{kl}\,,
\label{eq:iso-alg}
\end{equation}
where the Riemann tensor appears in the $[P,P]$ commutator and the $\mathfrak{so}(n)$ generators $M_{ij}$ should be suitably projected to the subalgebra $\mathfrak h$. Note that the superisometry algebra  $\mathfrak g=\mathfrak h\oplus\mathfrak p$ with $\mathfrak p$ spanned by $\{P_i\}$ and $\mathfrak h$ spanned by the suitably projected $\{M_{ij}\}$. Our strategy will be to derive the form of the Riemann tensor by casting the superisometry algebra of the space in question into this form. Note also that it is important that the $P_i$ are correctly normalized, i.e. $\mathrm{Tr}(P_iP_j)=-\frac{2}{R^2}\delta_{ij}$, with all generators anti-hermitian.

In the following we define $E_{ij}$ to be matrix with $1$ in position $(i,j)$ and zero elsewhere. We also define $A_{ij}=E_{ij}-E_{ji}$ and $S_{ij}=E_{ij}+E_{ji}$.

\subsection{$SLAG_3$}
$SLAG_3$ is the five-dimensional symmetric space $SU(3)/SO(3)$. The Lie algebra of $SU(3)$ consists of anti-hermitian traceless $3\times3$ matrices. The isometry algebra splits as $\mathfrak{su}(3)=\mathfrak{h}\oplus\mathfrak{p}$ where we take $\mathfrak{h}=\mathfrak{so}(3)$ to be spanned by the anti-symmetric real matrices
\begin{equation}
X_1=A_{12}\,,\quad X_2=A_{13}\,,\quad X_3=A_{23}\,,
%[X_1,X_2]=-X_3 etc.
\end{equation}
while $\mathfrak p$ is spanned by the imaginary symmetric matrices
\begin{equation}
Y_1=iS_{12}\,,\quad Y_2=i(E_{11}-E_{22})\,,\quad Y_3=-iS_{13}\,,\quad Y_4=-iS_{23}\,,\quad Y_5=\frac{i}{\sqrt3}(E_{11}+E_{22}-2E_{33})\,.
%
%[X_1,Y_1]=2Y_2
%[X_1,Y_3]=-Y_4
%[X_1,Y_4]=Y_3
%[X_1,Y_5]=0
%[X_2,Y_3]=-Y_2-\sqrt3Y_5
%[X_2,Y_4]=-Y_1
%[X_2,Y_5]=\sqrt3Y_3
\end{equation}
The realization in terms of generators $M_{ij}$ and $P_i$ in (\ref{eq:iso-alg}) looks like
\begin{equation}
X_1=2M_{12}-M_{34}\,,\quad
X_2=M_{14}+M_{23}-\sqrt3M_{35}\,,\quad
X_3=M_{13}-M_{24}-\sqrt3M_{45}\,,\qquad Y_i=P_i\,.
%
%[X_1,X_2]=-X_3 ok
%[X_1,Y_1]=2Y_2 ok
%[X_1,Y_3]=-Y_4 ok
%[X_1,Y_4]=Y_3 ok
%[X_1,Y_5]=0 ok
%[X_2,Y_3]=-Y_2-\sqrt3Y_5 ok
%[X_2,Y_4]=-Y_1 ok
%[X_2,Y_5]=\sqrt3Y_3 ok
\end{equation}
It is not hard to show that this indeed reproduces the correct commutation relations using (\ref{eq:iso-alg}). We can now read off the Riemann tensor from the $[P_i,P_j]$ commutator and we find the non-zero components
\begin{align}
%[Y_1,Y_2]=2X_1
%[Y_1,Y_3]=X_3
%[Y_1,Y_4]=X_2
%[Y_1,Y_5]=0
%[Y_2,Y_3]=X_2
%[Y_2,Y_4]=-X_3
%[Y_2,Y_5]=0
%[Y_3,Y_4]=-X_1
%[Y_3,Y_5]=-\sqrt3X_2
%[Y_4,Y_5]=-\sqrt3X_3
&R_{12}{}^{12}=-4\,,\quad
R_{13}{}^{13}=R_{14}{}^{14}=R_{23}{}^{23}=R_{24}{}^{24}=R_{34}{}^{34}=-1\,,\quad
R_{35}{}^{35}=R_{45}{}^{45}=-3\,,\quad
\nonumber\\
&R_{12}{}^{34}=R_{34}{}^{12}=2\,,\quad
R_{13}{}^{24}=R_{24}{}^{13}=1\,,\quad
R_{13}{}^{45}=R_{45}{}^{13}=\sqrt3\,,\quad
R_{14}{}^{23}=R_{23}{}^{14}=-1\,,\quad
\nonumber\\
&R_{14}{}^{35}=R_{35}{}^{14}=\sqrt3\,,\quad
R_{23}{}^{35}=R_{35}{}^{23}=\sqrt3\,,\quad
R_{24}{}^{45}=R_{45}{}^{24}=-\sqrt3\,.
\label{eq:SLAG3-R}
\end{align}

\subsection{$\mathbbm{CP}^3$}
$\mathbbm{CP}^3$ is the six-dimensional symmetric space $U(4)/(U(3)\times U(1))=SU(4)/S(U(3)\times U(1))$. The Lie algebra of $SU(4)$ consists of anti-hermitian traceless $4\times4$ matrices. The isometry algebra splits as $\mathfrak{su}(4)=\mathfrak h\oplus\mathfrak p$ where we take $\mathfrak{h}=\mathfrak{su}(3)\oplus\mathfrak u(1)$ to be spanned by the anti-hermitian traceless $3\times3$ block plus the $U(1)$ generator
\begin{align}
&X_i=i(E_{i+1,i+1}-\tfrac14\mathbbm1)\quad (i=1,2,3)\,,\quad X_4=A_{12}\,,\quad X_5=A_{13}\,,\quad X_6=A_{23}\,,\quad X_7=iS_{12}\,,\nonumber\\
&X_8=iS_{13}\,,\quad X_9=iS_{23}\,,
%[X_1,X_4]=-X_7
%[X_1,X_5]=0
%[X_1,X_6]=X_9
%[X_1,X_7]=X_4
%[X_1,X_8]=0
%[X_1,X_9]=-X_6
\end{align}
while $\mathfrak p$ is spanned by the remaining matrices
\begin{equation}
Y_1=A_{14}\,,\quad Y_2=iS_{14}\,,\quad Y_3=A_{24}\,,\quad Y_4=iS_{24}\,,\quad Y_5=A_{34}\,,\quad Y_6=iS_{34}\,.
%
%[X_1,Y_3]=Y_4
%[X_1,Y_4]=-Y_3
%[X_2,Y_5]=Y_6
%[X_2,Y_6]=-Y_5
%[X_3,Y_1]=-Y_2
%[X_3,Y_2]=Y_1
%[X_3,Y_3]=-Y_4
%[X_3,Y_4]=Y_3
%[X_3,Y_5]=-Y_6
%[X_3,Y_6]=Y_5
%[X_4,Y_1]=-Y_3
%[X_4,Y_2]=-Y_4
%[X_4,Y_3]=Y_1
%[X_4,Y_4]=Y_2
%[X_6,Y_3]=-Y_5
%[X_6,Y_4]=-Y_6
%[X_7,Y_1]=Y_4
%[X_7,Y_2]=-Y_3
\end{equation}
The realization in terms of generators $M_{ij}$ and $P_i$ in (\ref{eq:iso-alg}) looks like
\begin{align}
&X_1=M_{34}\,,\quad
X_2=M_{56}\,,\quad
X_3=-M_{12}-M_{34}-M_{56}\,,\quad
X_4=-M_{13}-M_{24}\,,
\nonumber\\
&
X_5=-M_{15}-M_{26}\,,\quad
X_6=-M_{35}-M_{46}\,,\quad
X_7=M_{14}-M_{23}\,,\quad
X_8=M_{16}-M_{25}\,,
\nonumber\\
&
X_9=M_{36}-M_{45}\,,\qquad
Y_i=P_i
\end{align}
We can now read off the Riemann tensor from the $[P_i,P_j]$ commutator and we find that it takes the simple form
\begin{equation}
%Y_1=A_{14}\,,\quad Y_2=iS_{14}\,,\quad Y_3=A_{24}\,,\quad Y_4=iS_{24}\,,\quad Y_5=A_{34}\,,\quad Y_6=iS_{34}\,.
%R_{12}^{12}=-4
%R_{12}^{34}=R_{12}^{56}=-2
%R_{13}^{13}=R_{13}^{24}=-1
%
R_{ab}{}^{cd}=-2(\delta_{[a}^c\delta_{b]}^d+J_{[a}{}^cJ_{b]}{}^d+J_{ab}J^{cd})\,,
\label{eq:CP3-R}
\end{equation}
were $J_{ab}$ is anti-symmetric with non-zero components given by $J_{12}=J_{34}=J_{56}=1$, i.e. the components of the K\"ahler form.

\subsection{$G_{\mathbbm R}^+(2,5)$}
$G_{\mathbbm R}^+(2,5)$ is the six-dimensional symmetric space $Sp(2)/U(2)$. The isometry algebra splits as $\mathfrak{sp}(2)=\mathfrak h\oplus\mathfrak p$ where we take $\mathfrak{h}=\mathfrak{u}(2)$ to be spanned by
\begin{equation}
X_1=i(E_{11}-E_{33})\,,\quad X_2=i(E_{22}-E_{44})\,,\quad X_3=A_{12}+A_{34}\,,\quad X_4=i(S_{12}-S_{34})\,,
\end{equation}
while $\mathfrak p$ is spanned by the remaining matrices
\begin{equation}
Y_1=A_{13}\,,\quad Y_2=iS_{13}\,,\quad Y_3=A_{24}\,,\quad Y_4=iS_{24}\,,\quad Y_5=\tfrac{1}{\sqrt2}(A_{14}+A_{23})\,,\quad Y_6=\tfrac{i}{\sqrt2}(S_{14}+S_{23})\,.
%
%[X_1,Y_1]=2Y_2
%[X_1,Y_2]=-2Y_1
%[X_1,Y_5]=Y_6
%[X_1,Y_6]=-Y_5
%[X_2,Y_5]=Y_6
%[X_3,Y_1]=-\sqrt2Y_5
%[X_3,Y_2]=-\sqrt2Y_6
%[X_3,Y_3]=\sqrt2Y_5
%[X_3,Y_4]=\sqrt2Y_6
%[X_4,Y_1]=\sqrt2Y_6
%[X_4,Y_2]=-\sqrtY_5
%[X_4,Y_3]=\sqrt2Y_6
\end{equation}
The realization in terms of generators $M_{ij}$ and $P_i$ in (\ref{eq:iso-alg}) looks like
\begin{align}
&X_1=2M_{12}+M_{56}\,,\quad
X_2=2M_{34}+M_{56}\,,\quad
X_3=-\sqrt2(M_{15}+M_{26}-M_{35}-M_{46})\,,
\nonumber\\
&X_4=\sqrt2(M_{16}-M_{25}+M_{36}-M_{45})\,,\qquad
Y_i=P_i\,.
%
%[X_1,X_2]=0
%[X_1,X_3]=X_4
%[X_1,X_4]=-X_3
%[X_2,X_3]=-X_4
\end{align}
We can now read off the Riemann tensor from the $[P_i,P_j]$ commutator and we find the non-zero components
\begin{align}
&R_{12}^{12}=R_{34}^{34}=-4\,,\quad
R_{12}^{56}=R_{34}^{56}=R_{56}^{12}=R_{56}^{34}=R_{56}^{56}=-2\,,\quad
R_{15}^{15}=R_{15}^{26}=-R_{15}^{35}=-R_{15}^{46}=-1\,,
\nonumber\\
&
R_{16}^{16}=-R_{16}^{25}=R_{16}^{36}=-R_{16}^{45}=-1\,,\quad
R_{25}^{16}=-R_{25}^{25}=R_{25}^{36}=-R_{25}^{45}=1\,,
\nonumber\\
&
R_{26}^{15}=R_{26}^{26}=-R_{26}^{35}=-R_{26}^{46}=-1\,,\quad
R_{35}^{15}=R_{35}^{26}=-R_{35}^{35}=-R_{35}^{46}=1\,,
\nonumber\\
&
R_{36}^{16}=-R_{36}^{25}=R_{36}^{36}=-R_{36}^{45}=-1\,,\quad
R_{45}^{16}=-R_{45}^{25}=R_{45}^{36}=-R_{45}^{45}=1\,,
\nonumber\\
&
R_{46}^{15}=R_{46}^{26}=-R_{46}^{35}=-R_{46}^{46}=1\,.
\label{eq:GR25-R}
\end{align}

%%%%%%%%%%%%%%%%%%%%%%

\end{document}